\documentstyle[12pt]{article}
\newcommand{\beq}{\begin{equation}}
\newcommand{\eeq}{\end{equation}}
\newcommand{\bef}{\begin{figure}}
\newcommand{\enf}{\end{figure}}
\newcommand{\bdis}{\begin{displaymath}}
\newcommand{\edis}{\end{displaymath}}

\title{Mixing in a Meandering Jet:\\ a Markovian Approximation}
\author{ M. Cencini$^1$, G. Lacorata$^{2,\ast}$, 
A. Vulpiani$^1$ and E. Zambianchi$^3$ }
\begin{document}

\maketitle
\centerline{$^{1}$  Dipartimento di Fisica, Universit\`{a} di  Roma 
"la Sapienza"}
\centerline{Piazzale Aldo Moro 5, I-00185 Roma, Italy and}
\centerline{Istituto Nazionale Fisica della Materia, Unit\`a di Roma}

\centerline{$^2$ Dipartimento di Fisica, Universit\`{a} 
dell' Aquila }
\centerline{Via Vetoio 1, I-67010 Coppito, L'Aquila, Italy}

\centerline{$^{3}$  Istituto di Meteorologia e Oceanografia Istituto}
\centerline{Universitario Navale, Corso Umberto I 174,
I-80138 Napoli, Italy.}

\vspace{.5cm}
\centerline{$^{\ast}$ and Istituto di Fisica dell'Atmosfera - CNR}
\centerline{P.le Luigi Sturzo 31, I-00144 Roma, Italy}

\date{ }
\medskip

\begin{abstract}
In this paper we investigate mixing and transport in correspondence of a 
meandering jet. The large-scale flow field is a kinematically assigned 
streamfunction. Two basic mixing mechanisms are considered, first separately
and then combined together: deterministic chaotic advection, induced by
a time dependence of the flow, and turbulent diffusion, described by means
of a stochastic model for particle motion.

Rather than looking at the details of particle trajectories, fluid exchange
is studied in terms of markovian approximations. The two-dimensional
physical space accessible to fluid particles is subdivided into regions
characterized by different Lagrangian behaviours. From the observed transitions
between regions it is possible to derive a number of relevant quantities 
characterizing transport and mixing in the studied flow regime, such as
residence times, meridional mixing, correlation functions.
These estimated quantities are compared with the corresponding ones
resulting from the actual simulations.
The outcome of the comparison suggests the possibility of describing
in a satisfactory way at least some of the mixing properties ot the
system through the very simplified approach of a first order markovian
approximation, whereas other properties exhibit memory
patterns of higher order.
\end{abstract}

Key words: Gulf Stream, mixing, chaotic advection, turbulent diffusion,
Markov process.
 
\vspace{.5cm}

PACS numbers: 92.10.Lq, 92.10.Fj, 05.45.+b

\renewcommand{\baselinestretch}{2.} 

\section{Introduction}

Western boundary current extensions typically exhibit 
a meandering jet-like flow pattern.
The most renowned example of this is given
by the meanders of the Gulf Stream extension, which have been investigated
in their variability by means of both hydrographic/currentmetric
and remotely sensed data 
(see, e.g., Watts, 1983, for a survey of earlier studies; Halliwell 
and Mooers, 1983; Vazquez and Watts, 1985;
Cornillon et al., 1986;
Tracey and Watts, 1986; 
Kontoyiannis and Watts, 1994; Lee, 1994).

These strong currents often separate very different regions of the
oceans, characterized by water masses which are quite different in terms
of their physical and biogeochemical characteristics. Consequently,
they are associated with very sharp and localized property gradients;
this makes the study of mixing processes across them 
particularly relevant also for interdisciplinary investigations.
This is the case of the Gulf Stream (Bower et al., 1985; 
Wishner and Allison, 1986;
Bower and Lozier, 1994), 
of the Kuroshio and of the Brazil/Malvinas current (Backus, 1986).

The mixing properties of passive tracers across meandering jets
has been investigated in the recent past by a number of authors, 
following essentially two different approaches.
The first one is that of dynamical models, where the flow is 
produced by integrating the equations of motion, time dependence is
typically produced by (barotropic or baroclinic) instability processes,
and dissipation is present (e.g., Yang, 1996). These
models account for several mechanisms acting in mixing in the real
ocean, even if sorting out single processes of interest may be
sometimes tricky.

A second and simpler approach, the one followed in this paper,
is that of kinematic models (Bower, 1991; Samelson, 1992; 
Dutkiewicz et al., 1993 -- hereafter respectively referred to
as B91, S92, DGO93 -- Duan and Wiggins, 1996; for slightly
different kinds of flows see also Lacorata et al., 1995).
In such models the large-scale velocity field is represented
by an assigned flow whose spatial and temporal characteristics mimic
those observed in the ocean. However, the flow field may not be
dynamically consistent in the sense of being a solution of the
equations of motion, or of conserving, e.g., potential vorticity.
Despite their somehow artificial character, these simplified models
enable to focus on very basic mixing mechanisms.

The paper B91
represents a first attempt at understanding
particle exchange in a two-dimensional meandering jet steadily propagating
eastward. 

The large-scale flow proposed in B91 has been utilized as a background
field in further works where mixing is separately enhanced by two
different transport mechanisms. S92 considers 
a modification of the B91 flow field where fluid exchange is induced by 
chaotic advection generated by a flow time dependence.
The basic flow is made time-dependent in three different fashions:
the superposition of a time-dependent meridional velocity;
that of a propagating plane wave;
a time oscillation of the meander amplitude, which is the case we 
further investigate in this paper.

The Melnikov method (see Lichtenberg and Lieberman 1992, LL92 hereafter) 
is used in S92 to explore the chaotic behaviour
around the separatrices of the original B91 flow 
when a time dependence is added. 
One of the results of this investigation is that while mixing 
occurs between adjacent regions, over a broad range of the 
meander oscillation frequencies, it does not easily take place across 
the jet, i.e. from recirculations south of the jet to recirculations
north of it. This is inherently due to the oscillation pattern of the
large-scale velocity, and we will discuss this in detail further in
this paper.

Particle exchange in the same B91 flow is achieved by DGO93 
by superimposing to the original 
time-independent basic flow a stochastic term which 
describes mesoscale turbulent diffusion in the upper ocean. 
The focus of that paper is on the exchange among recirculations 
and jet core and viceversa, and on the homogenization 
processes in the recirculation. The numerical experiments presented in DGO93
are carried out for quite short integration times, which do not allow
for exploring the mixing across the jet.

Since in the real ocean the two above mixing mechanisms, i.e. chaotic 
advection and turbulent diffusion, are simultaneously present, in this 
paper we investigate
how particle exchange varies through the progression from periodic 
to stochastic disturbances, revisiting and putting together
the mixing processes studied by S92 and DGO93. 

This is done by looking at particle statistics obtained by numerical
computation of the trajectories of a large number 
of particles (or equivalently, since our
system is ergodic, following one particle 
for a very long time) in three different flows: one equivalent to S92,
in which mixing is induced by chaotic advection; one equivalent to DGO93,
where it is due to turbulent diffusion, and a combination of them.

Dispersion processes in a flow field can be quantitatively characterized,
in the Lagrangian description, in terms of different quantities, such
as, e.g.,
the Lyapunov exponent $\lambda$ (Benettin et al. 1980)
and the diffusion coefficients $D_{ij}$ (LL92).

However the above indicators, even if mathematically well defined, 
can be rather unrelevant for many purposes.
The Lyapunov exponent is the inverse of a characteristic time $t_{L}$,
related to the exponential growth of the distance between two
trajectories initially very close;
however, other characteristic time scales may appear and result as
relevant in the description of a system, such as
those involved in the correlation functions
and in the mixing phenomena. It is worth stressing that there 
is not a clear relationship, if any, among these times and  $t_{L}$. 

Also the use of the diffusion coefficients can have severe limitations;
sometimes the $D_{ij}$ are not able to take into account the basic 
mechanisms of the spreading and mixing (Artale et al., 1997).
Our western boundary current extension 
system has essentially a periodic structure in the zonal direction
It is thus possible to define and (numerically) compute the diffusion 
coefficients.
They are related to the asymptotic behaviour, i.e. long times and large
spatial scales, of a cloud of test particles.
On the other hand,
if one is interested in the meridional 
mixing, which takes place over finite time scales,
the diffusion coefficients may not be very useful.
In such situations it is then worthwhile
to look for alternative methods of describing
mixing processes, as was done by
Artale et al. (1997) looking at dispersion in closed basins
or by Buffoni et al. (1997), who employ
exit times for the characterization of transport in basins with complicated
geometry.

\vspace{4mm}

Our investigation is carried out with a non conventional approach, 
in a geophysical contest, as we try to analyze 
the system from the standpoint of the approximation 
in terms of markovian processes (Cecconi and Vulpiani 1995, Kluiving et al
1992, Nicolis et al. 1997, Fraedrich and M\"uller 1983 and Fraedrich 1988).

We start from the consideration that the flow field we want to characterize
in terms of fluid transport can be subdivided into regions corresponding 
to different lagrangian behaviours: ballistic fly in the meandering jet core,
trapping inside recirculations, retrograde motion in the far field. As an
obvious consequence,
we introduce a partition of the two-dimensional physical space
accessible to fluid particles and divide it into these disjoint regions
selected in a natural way by the dynamics.
At this point, one can stydy the transition of fluid particles
between different regions as a discrete stochastic process generated
by the dynamics itself.

In this paper we study the statistical properties of this stochastic 
process and we compare it with an approximation in terms of Markov 
chains.
For some fluid exchange properties --- such as, e.g., the probability
distribution of the particle exit times from the jet or from
the neigbouring recirculation regions --- the effects of the two
different mixing mechanisms and the results of the markovian 
approximation are very similar.
Other properties, such as the meridional mixing across the jet,
do not show such an obvious possibility to be described 
in terms of markovian simple processes.
However, for those properties the markovian description 
is seen to be relatively more accurate in the case when 
chaotic advection and turbulent diffusion are simultaneosly 
present.

The comparison between the results of the numerical simulations 
and those computed in the markovian approximation
allows for a deeper understanding of the transport and 
mixing mechanisms.

In section 2 we introduce the kinematic model for the 
flow field correspondent to the Gulf
Stream flow and both models for chaotic advection 
and turbulent diffusion.
Section 3 is devoted to the description of the markovian 
approximation.
In Section 4 we discuss the numerical results and the comparison of 
the true dynamics with the markovian approximation.
Section 5 contains some discussion and conclusions.
The Appendix summarizes some basic properties of Markov chains.

\section{The flow field}

The large-scale flow in its basic form,
representing the velocity field in correspondence of a meandering
jet, is the same introduced in B91 and further discussed
in S92; in a fixed reference frame the streamfunction is given by

\beq
\psi(x',y',t) \,=\,
\psi_0 \left[1-tanh \frac{y'-A cos\, \kappa(x'-c_x t)}
{\lambda(1+k^2A^2sin^2 \kappa(x'-c_x t))^{1/2}} \right] 
\label{eq:psi}
\eeq

where ${\psi_0}$ represents half of the total transport,
$A$, $k$ and $c_x$ the amplitude, wavenumber and phase speed of the 
pattern.   
A change of coordinates into a reference frame moving eastward with a 
velocity coinciding with the phase speed $c_x$, and a successive 
nondimensionalization, yield a streamfunction as follows:

\beq
\phi \,=\,-tanh\left[ \frac{y-Bcos\, kx}{(1+k^2B^2sin^2kx)^{1/2}} \right] 
+ cy
\label{eq:phi}
\eeq

The relationship between variables in (\ref{eq:psi})  and 
(\ref{eq:phi}) is given by (see S92):

$$
x \equiv \frac{x'-c_x t}{\lambda},
\;\;
y \equiv y'/\lambda,
\;\;
B \equiv A/\lambda
$$

$$
\phi \equiv \frac{\psi}{\psi_0} + cy,
\;\;
c \equiv \frac{c_x L}{\psi_0},
\;\;
\kappa \equiv k \lambda
$$

The evolution of the tracer particles is given by:
\beq 
\frac{{\rm d}x}{{\rm d}t}=-\frac{\partial \phi}{\partial y}\,,\;\;
\frac{{\rm d}y}{{\rm d}t}=\frac{\partial \phi}{\partial x}\,.
\label{eq:hamil}
\eeq

The natural distance unit for our system is given by the jetwidth
$\lambda$, set to 40 Km (B91, S92). The basic flow configuration is very
similar to case (b) of B91: $B$ was chosen as $1.2$; $c$ as $0.12$.
The only sensible difference is the value assigned to $L$, i.e. 
the meander wavelength, which was set as $7.5$, as will be discussed
below.

In fig. 1 we show the stationary velocity vector 
field in the moving frame: the field
is evidently divided into three very different flow regions (see also
B91, S92, DGO93): the central, eastward moving jet stream, recirculation
regions north and south of it, and a far field; the far field, 
given our choice of parameters, appears to be moving westward at a phase
speed of $-c_x \equiv -0.12$.
This intrinsic self subdivision of the flow field will result
crucial for building a partition of the possible states available
to our test particles, which will be investigated in markovian terms.

Chaotic advection may be induced in a two-dimensional flow field by
introducing a time dependence (see e.g. Crisanti et al., 1991).
This is simply achieved by adding to
the basic steady flow some tipically small perturbation which varies in
time. Among three basic mechanisms discussed by S92, we chose here a 
time-dependent oscillation of the meander amplitude:

\beq
B(t) \,=\, B_0 + \epsilon \cdot \cos(\omega t + \theta)
\label{eq:pulsaz}
\eeq

In (\ref{eq:pulsaz}) we set $B_0 = 1.2$, $\epsilon = 0.3$, $\omega
= 0.4$ and $\theta = \pi/2$. These choices, as well as that for $L$,
are motivated mainly by the results of the observations by Kontoyiannis
and Watts (1994) and of the numerical simulations by Dimas and
Triantafylou (1995). Namely, the most unstable waves produced in the latter
work compare very well with the observations of the former, which show
wavelengths of 260 Km, periods of $\sim$ 8 days, e-folding space and
time scales of 250 Km and 3 days respectively.
In our case, since the downstream speed was set to $1\, m/s$,
our e-folding time scale would correspond, in dimensional units,
to approximately 3.5 days. The flow field resulting from the time
dependent version of (\ref{eq:phi}) is shown in fig. 2 for three
different subsequent time snapshots $t=0.\,\,(T/2)$ (fig. 2a), $t=T/4$ 
(fig. 2b) and $t=3T/4$ (fig. 2c).
Our system shows two different
separatrices with a spatial periodic structure (see fig. 2),
North and South of the jet.
At small $\epsilon$ one has chaotic motion around them but 
without meridional mixing.
In order to have a ``large scale chaos'', i.e. the possibility
that a test particle passes from North to South (and viceversa),
crossing the jet, one needs the {\it overlapping of the resonances}
(Chirikov 1979)  $\epsilon>\epsilon_{c}$. 
In our case $\epsilon>\epsilon_{c}$ and $\epsilon_{c}$ depends on $\omega$ 
(in fig. 3 we show $\epsilon_{c}$ vs $\omega$ for our system).

The lagrangian motion of a test particle
is formally a Hamiltonian system whose Hamiltonian is 
the stream function $\phi$.
If $\phi=\phi_{0}(x,y)+\delta \phi (x,y,t)$, where
$\delta \phi (x,y,t)=O(\epsilon)$ is a periodic function of $t$, 
there exists a well
known technique, due to Poincar\'e and Melnikov, which allows to 
prove whether the motion is chaotic (LL92). 
Basically if the steady part $\phi_{0}(x,y)$ of the stream
functions admits homoclinic (or heteroclinic) orbits, i.e.
separatrices, then the motion is usually chaotic in a small region around
the separatrices for small values of $\epsilon$.

In S92 the Melnikov integral has been computed explicitly 
for the $\phi_{0}$ and $\delta \phi$ that we have
used in this paper, and thus proved the existence of lagrangian chaos.
However, even if
the Melnikov method can determine if the lagrangian motion is
chaotic, it is not
powerful enough for the study 
of other interesting properties which will be the focus of
section 4.1.

Alternatively (or jointly), mixing in the flow field (\ref{eq:phi}) 
can be created by adding a turbulent diffusion term. This was done 
utilizing a stochastic model for particle motion belonging to the 
category of the so-called "random flight" models (e.g. Thomson, 1987),
which can be seen as simple examples of a more general class of stochastic
models which can be nonlinear and have arbitrary dimensions, described
by the generalized Langevin equations (Risken, 1989; for a review, see
Pope, 1994):
\beq
ds_i= h_i({\bf{s}},t)dt + g_{i,j}({\bf{s}},t) d\mu_j   
\;\;\;\;\;\;\;\;\;\;\;\;\;[i=1,\ldots,N]
\label{eq:lang}
\eeq
where ${\bf s}=(s_{1},\ldots,s_{N})$ are N stochastic variables 
which, in our context, are the turbulent velocity fluctuations,
$\mu_i$  is a random process with independent
increments, and $h_{i}$ and 
$g_{i,j}$ are continuous 
functions.
A general, remarkable characteristic of these models is their markovian
nature, which obviously has a particular interest for this investigation.
The theoretical motivation for the choice of markovian models to describe
mesoscale ocean turbulence has been thoroughly discussed in Zambianchi
and Griffa (1994a), Griffa (1996) and Lacorata et al. (1996); 
it is worth adding that this 
particular model has proved to accurately represent upper ocean 
turbulence in regions characterized by homogeneity and stationarity
(see Zambianchi and Griffa, 1994b; Griffa et al., 1995; Bauer et al.,
1997; but also the results of numerical simulations by 
Verron and Nguyen, 1989; Yeung and Pope, 1989; Davis, 1991),
and is easily extended to more complex situations (van Dop
et al., 1985; Thomson, 1986).

In our simulations, a turbulent velocity 
$\delta {\bf u}^{(T)}({\bf x}, t)$
is added to the 
large-scale velocity field
$ {\bf u}^{(M)}({\bf x}, t)$ 
resulting from the streamfunction (\ref{eq:phi}).
The resulting equation for the particle trajectory is:
\beq   
{{d {\bf x}} \over {d t}}= {\bf u}({\bf x}, t)
\label{eq:l1}
\eeq
where 
{\bf u}({\bf x}, t) is given by
\beq
 {\bf u}({\bf x}, t)= {\bf u}^{(M)}({\bf x}, t)+ 
\delta {\bf u}^{(T)}({\bf x}, t).
\label{eq:l2}
\eeq
Our model assumes $\delta {\bf u}^{(T)}({\bf x}, t)$ as a Gaussian process  
with zero mean and correlation:
\beq
<    \delta u_i^{(T)}({\bf x}, t) \delta u_j^{(T)}({\bf x'}, t')>=
 2 \, \sigma^2 \delta_{ij} \delta({\bf x}-{\bf x'}) e^{-|t-t'|/\tau}.
\label{eq:gauss}
\eeq
With this choice, $\delta {\bf u}^{(T)}({\bf x},t)$ 
is a linear in time markovian process:
the turbulent field is entirely described 
in terms of two parameters: the variance 
of the small scale velocity $\sigma^2$ and the e-folding time scale of the 
velocity 
autocorrelation function, i.e. its typical correlation time scale $\tau$,
which represents the time step of the markovian process.
The interdependence among smaller and larger time scales of
the lagrangian motion will be investigated in the following chapters.

\vspace{1cm} 
  
\section{The Markovian approximation}

The idea to use stochastic processes to study (and describe) 
chaotic behaviour is rather old (Chirikov 1979, Benettin 1984). 
One of the most relevant and successful approach
is the symbolic dynamic, which allows to give a detailed 
description of the statistical properties of a chaotic system in terms
of a suitable discrete stochastic process (Beck and Sch\"ogl 1993).

Given a discrete dynamical system:
\beq
{\bf x}_{t}=S^{t}  {\bf x}_{0}\,,
\label{eq:evol}
\eeq
one can introduce a partition $\cal{A}$ dividing the phase 
space in $A_{1},A_{2},\cdots,A_{N}$ disjoint sets
(with $A_{i} \cap A_{j}=0$ if $i \neq j$). 
From each initial condition one has a trajectory:
\beq
{\bf x}_{0},{\bf x}_{1},\cdots,{\bf x}_{n},\cdots\; 
\label{eq:traj}
\eeq
The point ${\bf x}_{0} \in A_{i_{0}}$ will select the integer $i_0$, 
the next one ${\bf x}_{1} \in A_{i_1}$ the integer $i_1$ 
and so on. Therefore for any initial condition 
${\bf x}_{0}$ we have a certain symbol sequence:
\beq
{\bf x}_{0} \rightarrow (i_{0},i_{1},\cdots,i_{n},\cdots)\,.
\label{eq:symb}
\eeq

Now the study of the {\it coarse grained\/} properties of the chaotic 
trajectories is reduced to the statistical features of the 
discrete stochastic process $(i_{0},i_{1},\cdots,i_{n},\cdots)$.
A useful and important characterization of the properties of
symbolic sequences is the Kolmogorov-Sinai entropy, defined by:
\begin{eqnarray}
h_{KS}&=&\lim_{n\rightarrow \infty} (H_{n+1}-H_{n})\,,
\label{eq:entroKS}
\end{eqnarray}
with
\begin{eqnarray}
H_{n}&=& \sup_{\{\cal{A}\}} [- \sum_{C_{n}} P(C_n) \ln P(C_n)]\,,
\label{eq:sup}
\end{eqnarray}
and
\begin{eqnarray}
C_{n}&=&(i_{0},i_{1},\cdots,i_{n-1})\,,
\label{eq:seq}
\end{eqnarray}
where  $P(C_n)$ is the probability of the sequence $C_{n}$
and $\{\cal{A}\}$ is the set of all possible partitions.

We note that, from a theoretical point of view, 
the $\sup$ in (\ref{eq:sup}) hides a very subtle point:
sometimes there exist a particular partition, 
called {\it generating partition\/}, for which one has 
automatically the $\sup$.
A partition is generating if the infinite symbol sequence
 $i_{0},i_{1},\cdots,i_{n},\cdots$ uniquely determines the initial value 
${\bf x}_{0}$.
Unfortunately it is not trivial at all to know if a system possesses
a generating partition. Moreover in practical applications, even if
the system admits a generating partition, it is extremely hard
to find it (see Beck and Sch\"ogl 1993 for more detail).

A part the difficulties for the choice of a suitable partition, 
the stochastic process given by the symbol dynamics with 
a given partition can have rather nontrivial features.
Of course the optimal case is when the symbolic stochastic 
process is a Markov chain, i.e. the probability to be in the cell
$A_i$ at time $t$  depend only on the cell at time $t-1$.
In this case is possible to derive all the statistical properties 
(e.g. entropy and correlation functions) from the transition matrix 
$W_{ij}$ whose elements are the probabilities to have the system 
in the cell $A_{j}$ at time $t$ if at time $t-1$ the system is 
in the cell $A_{i}$. See the Appendix for a summary of the properties of
Markov chains.

Usually a $k$-order markovian process (i.e. one in which
the probability to be in the 
cell $A_{j}$ at time $t$ depends only on the preceding $k$ steps
$t-1,t-2,\cdots,t-k$), with large $k$, is necessary to mimic, 
with a good accuracy,
a chaotic system. 
In particular by means of the quantities defined
in eqs. (\ref{eq:entroKS},\ref{eq:sup}) and of consideration of 
information theory (see Khinchin 1957) we are able to estimate the order 
of the Markov process necessary to reproduce the statistics of 
the process $(i_0,i_1,\cdots,i_n,\cdots)$ generated by the dynamics.
In fact it is possible to demonstrate (Khinchin 1957) that, defining
\beq
h_n=H_n-H_{n-1}
\label{eq:block}
\eeq
with $H_{n-1}$ given by (\ref{eq:sup}), if the 
process is a Markov process of order $k$ then 
$h_n=h_{KS}$ for each $n \geq k+1$.
In the next section we apply this method, in our case, to give an estimate
of the order $k$.
 
However we observe that usually it is relatively easy, 
using a markovian approximation of order $k\leq 4 \div 5$, to find a 
reasonable agreement for the K-S entropy (\ref{eq:entroKS}) 
(or for the Lyapunov exponent). On the other hand for different 
properties, such as the correlation functions, it is necessary to 
use Markov processes of rather large order ($k \sim 8 \div 10$) 
just for a fair agreement.

The issue to mimic a low dimensional dynamical system in terms 
of a Markov process of a certain order is surely an interesting 
aspect in the field of chaotic dynamics but, in our opinion,
with a weak relevance for many practical purposes in geophysics since 
one needs a very large statistics for the computation of the 
transition probabilities.
Therefore we shall restrict our analysis to the simplest case
of the approximation in terms of a Markov chain, i.e. a 
first order process.
This practical approach has been successfully used in the study
of certain problems related to the dynamical properties 
of small astronomical bodies such as comets (see Rickmann and Froeshl\'e,
1979 and Levinson 1991) and for the interpretation of geophysical
phenomena (Fraedrich, 1988, Fraedrich and M\"uller, 1983).

\vspace{4mm}

We shall now explain how we proceeded to express the behaviour of our
Gulf Stream-like system in terms of symbolic cynamics.
First we reduced the ordinary differential equation 
(\ref{eq:hamil}) obtained by the stream function (\ref{eq:phi}) 
to a discrete in time dynamical system, which was accomplished 
building the Poincar\`e map associated with
(\ref{eq:hamil}).
This represents an assessment of the possibility to write
${\bf x}_{n+1}$ in terms of ${\bf x}_{n}$ -- defined as
${\bf x}_{n}\!=\!{\bf x}(t\!=\!\!nT)$ --:
\beq
{\bf x}_{n+1}={\bf F}[{\bf x}_{n}]\,.
\label{eq:poinc}
\eeq
Writing down an 
explicit expression for 
${\bf F}[{\bf x}_{n}]$
may be in general non trivial; however, it is worth noticing
the importance the existence of such a relationship.

We have now to decide a suitable choice of the phase space partition. 
Differently from the case of the orbital evolution of comets, 
where the cells are rectangular regions of the phase space, 
for our system we shall adopt a partition 
whose cells have curved frontiers.
Considering the streamline pattern of our flow field (fig. 2), the
structure of the physical space accessible to fluid particles
suggests an obvious, natural choice for the partition:
a particle will find itself in state 1 when it is
inside the jet core (open trajectories); states  2 and 3
correspond to trapping in the Northern and Southern recirculations respectively 
(closed trajectories) and states 
4 and 5 to the far field, i.e. far from the jet
(backward open trajectories).

This partition turns out to be particularly appropriate
to describe some important mixing properties 
in our system, such as:
\begin{itemize}
\item (a)
The residence times of particles in the trapping recirculations or inside
the jet, which in the language of Markov chain (see Appendix)
correspond to
the first exit times from state $i$
($i=1,\cdots,5$);
\item (b)
The meridional mixing times (MMT), i.e. the time it takes to a particle
to enter the Northern (Southern) recirculation starting from the 
Southern (Northern) one, i.e. the time of first 
passage from cell 2 (3) to cell 3 (2);
\item (c)
The correlation function for a variable 
$\chi_{i}(n)$ which indicates
if a determined state $i$ is visited at time $n$ (see below, eqs
\ref{eq:chi} and  \ref{eq:corr}, and Appendix).
\end{itemize}
Because of the system symmetries,
states 2 and 3 possess the same statistical 
properties and so do states 4 and 5;
in particular, the following equalities hold:
\begin{equation}
W_{12}=W_{13}\,,\; W_{23}=W_{32}\,,\;W_{21}=W_{31}\,,\; W_{22}=W_{33},
\;\;{\mbox {and  so  on}}
\label{eq:simmetries}
\eeq
  
Let us now describe how 
to compute statistics for the quantities 
(a,b,c).
First of all, we can compute from a long trajectory 
${\bf x}_{0},{\bf x}_{1},\cdots,{\bf x}_{n}$ 
($n\gg 1$) the transition probabilities:
\beq
W_{ij}=\lim_{n\rightarrow \infty} \frac{N_{n}(i,j)}{N_{n}(i)}
\label{eq:matrix}
\eeq
where $N_{n}(i)$ is the number of times that, along the trajectory, 
${\bf x}_{t}$ ($t<n$) visits the cell $A_{i}$
and $N_{n}(i,j)$ is the number of times
that ${\bf x}_{t} \in A_{i}$ and ${\bf x}_{t+1} \in A_{j}$.
In table 1 we report the elements of the matrix. 

Notice that, for each $i$ :
\beq
\sum_{j}W_{ij}=1
\label{AA}
\eeq
We can express the probabilities 
$\{P_{i}\}$ in terms of the matrix $\{W_{ij}\}$:
\beq
P_{i}=\sum_{j}P_{j}W_{ji}
\label{BB}
\eeq
Let us stress that the eqs. (\ref{AA}, \ref{BB})  hold even if
the process is not a Markov chain.  
Under the assumption (approximation) that the symbolic stochastic 
process generated by our deterministic chaotic model
is a Markov chain one can derive (see Appendix) 
the probability of the first exit times from state $i$ in $n$ steps:
\beq
P_{i}(n)=\left[ \frac{(1-W_{ii})}{W_{ii}}\right] (W_{ii})^{n}\,,
\label{eq:residence}
\eeq
which is the statistics of residence times in state $i$.
A slightly more complicated computation gives
the probabilities $f_{ij}(n)$ of first passage from state $i$ 
to state $j$ in $n$ steps:
\beq
f_{ij}(n)=(W^{n})_{ij}-\sum_{k=1}^{n-1} f_{ij}(n-k) (W^{k})_{jj}\,,
\label{eq:mixing}
\eeq
i.e. the statistics of the MMT.
For the normalized correlation function $C_{i}(n)$ of the variable 
$\chi_{i}(n)$ defined as:
\beq
\chi_{i}(n)=\left\{
\begin{array} {c}
1   \;\;\;   {\mbox {if}} \;{\bf x}_{n} \in A_{i} \\
0   \;\;\;   {\mbox {otherwise}}
\end{array}
\right.
\label{eq:chi}
\eeq
we have:
\beq
C_{i}(n)=((W^{n})_{ii}-P_{i})/(1-P_{i})\,.
\label{eq:corr}
\eeq

The Kolmogorov-Sinai entropy for the Markov chain is nothing but the 
Shannon entropy for a Markov chain:
\beq
h_{KS}=h_{S}=-\sum_{i,j} P_{i} W_{ij} \ln W_{ij}.
\label{eq:entroSh}
\eeq
Note that for a Markov process $h_n=H_n-H_{n-1}=h_{KS}$ for $n=2$
(see above).
Since the discrete time system is obtained looking it at times
$0,T,2T,\cdots$ the Lyapunov exponent $\lambda$ of the original system 
has to be compared with:
\beq
\lambda_{M}=\frac{h_{S}}{T}.
\eeq
This last equation is easily understood noticing that $h_{KS}$ gives
the degree of information per step produced by the process, that for a
chaotic system in two dimensions corresponds to the Lyapunov exponent 
apart from a time rescaling.

\section{Numerical results}

We now discuss the numerical results for the
models introduced in section 2 and their comparison with the
markovian approximation illustrated in section 3.

\subsection{Mixing induced by chaotic advection}
We first consider the deterministic model with
the parameter $B$ of the stream function eq. (\ref{eq:phi}) varying
periodically in time according to eq. (\ref{eq:pulsaz}) 
with the parameters  $B_0=1.2$, $\epsilon=0.3$, 
$\omega=0.4$, $\phi=\pi/2$ and $c=0.12$.
With this choice the system is chaotic and exhibits mixing
at large scale, i.e. North-South mixing occurs.
 
We show in fig. 4 the spreading at different 
times of a cloud of test particles. 
The domain is naturally defined from the basic cell 
that repeats itself creating a zonal
periodic structure of wavelength $L$; $x$ thus
varies in $[0,L]$, while $y$ in $[-4,4]$.
We fixed {\it a posteriori} these bounds for $y$ since  
for our choice of parameters no particles reach the far field, and
no trajectories attains values in $|y|$ 
larger than $|4|$
(even though, in general, we
expect low but non zero frequencies for these  
states, see also S92).

In general, whether a North-South mixing happens or not  
depends sensibly on the values of
$\epsilon$ and $\omega$.
Typically the system reveals a strong preference to have a long 
residence times in 
the Northern or the Southern half of the domain
with respect to the jet core. 
This peculiar feature will play an important 
role in the comparison with the markovian approximation.

The transition matrix elements $W_{ij}$ and 
the visit probabilities $\{P_i\}$ are computed by 
means of eq. (\ref{eq:matrix}), looking at ${\bf x}$ every
period i.e. for $t=T,2T,\cdots$ where $T=2 \pi/\omega$, see table 1 and 2. 
At a first glance, we can see that the requested symmetry properties 
are respected (eq. \ref{eq:simmetries}).

Now in order to test whether the system is well approximated by a first 
order Markov process we compute exit times, correlation
functions, meridional mixing times and Lyapunov exponent  
from the actual dynamics and compare them with the 
markovian predictions.

In fig. 5 we show the first exit time probability distributions for
the states 1, 2 (3) and the
corresponding markovian predictions.
After stressing that the straight lines of fig. 5   
are not to be confused with best-fit curves,
 we see that the agreement is good over a certain range 
both for state 1 and 2 (3). 
The agreement between the markovian predictions
and the actual results is rather poor for small and very
large exit times.
The above behaviour shows that the markovian 
approximation cannot hold at small times since the details
of the dynamics are strongly relevant.
In the same way non trivial long time correlations cannot
be accounted for in terms of a first order markovian process.

In fig. 6 we can see how the correlation functions of $\chi_i$ 
(see eqs. \ref{eq:chi}-\ref{eq:corr}) for the state 1 and 2 (3) are
just in vague agreement with the corresponding 
correlations obtained from the markovian process.  
The trajectories in the recirculations (i.e. states 2 and 3) 
appear to be  much more auto-correlated than those in the jet.
Therefore we deduce that the system, although chaotic, 
has a strong memory as to which
half (North or South) of the spatial domain it is visiting. 
So the typical evolution is
a ''rebound game'' between state 2 and the South half of the jet during a 
certain time interval; then it crosses the jet core and performs again
the same pattern between state 3 and the North half of the jet, until
it jumps back; and so on.  

This is strikingly evident looking at the distribution probabilities 
of the meridional mixing times, in fact in this case the markovian 
approximation completely fails (fig. 7).
 
The markovian approximation leads to a clear disagreement in this 
case because it actually happens that when a tracer leaves a 
recirculation region, say in the South half, and goes into the jet, 
most of the time it returns back to some other Southern close 
orbit rather than passing through the jet barrier, 
while in the first order Markov approximation it is 
not possible to explain such a strong memory effect.  
This feature is a clear indication that higher order Markov processes
are necessary to describe the statistics produced by the dynamics.
This is shown in fig. 8, where transitions between states 1 and 2, and 
1 and 3, are compared: whereas in the first order markovian approximation (solid
lines) a test particle jumps very often from North to South, the
results of our numerical experiments show a stronger tendency for particles
to keep being confined either between states 1 and 2 or between 1 and 3.

In order to quantify the relevance of the memory effects we 
computed the block entropies $h_n$, defined in 
section 3 (eq. \ref{eq:block}) at varying $n$. 
In fig. 9 we can see that to obtain the convergence 
of the entropies we need at least of a Markov approximation
of order $6 \div 7$.

The Lyapunov exponent computed with a standard algorithm (see Benettin 
et al. 1980) is $\lambda=0.05$, the first order Markov 
approximation gives $\lambda_{M}=0.03$, while the extrapolation
with the asymptotic value $h=\lim_{n \rightarrow \infty} h_n$
gives $\tilde{\lambda}_{M}=0.03$.
The fact that $\tilde{\lambda}_{M} < \lambda$ is probably due to the fact
that the partition here used is not a generating one
(see section 3) however there exist a fair agreement between 
$\tilde{\lambda}_{M}$ and $\lambda$.
It is worth noticing that 
the above features  are fairly robust, and 
do not vary in a relevant way after weak changes of parameters.

\subsection{Mixing induced by turbulent diffusion}
Now we perform the same investigation 
discussed in the previous subsection setting $\epsilon=0$ 
and turning on turbulent diffusion which is described in terms
of a stochastic model for particle motion:
\beq
\frac{dx}{dt} \,=\,u + \eta_{1}\,,\;\;
\frac{dy}{dt} \,=\,v + \eta_{2}\,,
\label{eq:noise}
\eeq
where $u,\;v$ are given by the stream function (\ref{eq:phi}) and 
$\eta_i$ are zero mean gaussian stochastic processes with
$<\eta_i(t) \eta_j(t')>=\sigma^2 \delta_{ij} \exp[-|t-t'|/\tau]$.
The gaussian variable $\eta_i$ are generated 
by a Langevin equation (Chandrasekhar, 1943):
\beq
\frac{d\eta_{i}}{dt} \,=\, -\frac{\eta_{i}}{\tau} 
+\sqrt{ \frac{2\sigma^2}{\tau}} \zeta_{i}, \;\;\; i=1,2 
\label{lange}
\eeq
where the variables $\zeta_i$ are zero mean Gaussian noises 
with $<\zeta_i(t)\zeta_j(t')>= \delta_{ij} \delta(t-t')$. 
The numerical integration of the equations
has been performed by means of a stochastic 4-th order Runge-Kutta 
algorithm (Mannella et al., 1989). 

Now the motion is unbounded for the presence
of the isotropic diffusive terms, so we need a ``trick'' to limit the
dispersion along $y$ inside a domain as similar as possible to the previous
one. In this way we can hope to look for a transition matrix comparable
with that one obtained in the chaotic deterministic case
 and compare the two models.
    
Since the chaotic model does not fill the states 4 and 5, 
we impose that if a tracer enters a backward motion region it is
reflected back by changing the sign of the meridional 
turbulent velocity, being the stream lines in these regions 
almost parallel to the $x$ direction. 
Let us stress that this boundary condition practically does not 
affect the mixing processes between 
the jet and the recirculation regions, therefore
we simply need to follow only those 
branches of trajectory which fall within the domain of study and  
not the large scale diffusion motion far from the stream frame.
  
In the diffusive case we fix the values $\sigma=0.05$ and $\tau=T/4$ 
as representative of an observable situation (see DGO93) 
and compute again the transition matrix and 
stationary frequencies of the 5 states (see tables 1 and 2).

In fig. 10 we show the spreading of a cloud of test particles.
Notice that the artifact introduced to bound the dynamics in the $y$ 
direction does not produce any artificial accumulation of particles
near the boundary.

At first we notice that the elements of the transition matrix are 
close to those of the chaotic case. 
We have computed the transition probabilities over 
a time period $T$ as for the chaotic case so that we can reasonably 
compare the two cases.

Fig. 11 shows the probabilities of the first exit times of the states 
1, 2 and 3 and the relative markovian predictions. 
These distributions are very well 
approximated by the first order Markov process.

The correlation functions are shown in Fig. 12. 
The difference between the 
actual and the markovian curves 
is now smaller than in the chaotic case
because of the presence of diffusion that decreases 
sensibly the degree of memory.

The most relevant difference with respect to the chaotic model is a 
clear improvement in the agreement with the markovian approximation for
the distributions of the meridional mixing times, see fig. 13.
The memory effects 
are now smoothened by diffusion, and the transition rates are much
more representative of the actual dynamics.
Thus we can conclude that in the diffusive model
the markovian approximation works much better than in the chaotic one.
Looking at the block entropies $h_n$ (\ref{eq:block}) we have a clear 
indication that the process is of a lower order respect the 
chaotic case (compare fig. 14  with fig. 9) 

Just like
in the previous case we have investigated the behaviour of the system 
varying the parameters $\sigma$ and $\tau$.
We have observed that if we keep the quantity $\sigma^{2} \tau$ 
constant the system displays similar behaviours, even if
the extent of the agreement between simulation results and markovian
approximation slightly differs for different values of the
turbulence parameters; this can be 
understood if we recognize that $\sigma^{2} \tau$ 
corresponds to the diffusion coefficient (see, e.g., Zambianchi and 
Griffa, 1994a). It has been shown that varying $\sigma$ and
$\tau$ even if keeping the diffusion coefficient constant can indeed
affect the quantitative estimates of dispersion in cases characterized
by inhomogeneity and/or nonstationariety (see, again, Zambianchi and
Griffa, 1994a). On the other hand, the qualitative functional 
behaviour of the dispersion processes has been seen to be influenced
very little by such changes in the parameters of turbulence (Lacorata
et al., 1996).

\subsection {Mixing jointly induced by chaotic advection and turbulent 
diffusion}

A detailed analysis of Lagrangian data from the ocean
aimed at determining contemporary presence and relative importance
of chaotic and turbulent mixing is at present still lacking,
as it would imply
the evaluation of both one- and two-particle statistics parameters,
which has been done only in the context of purely diffusive particle
exchange investigations (see, e.g., Poulain and Niiler, 1989).
However, in the real ocean we expect both the above mixing mechanisms,
discussed in sect. 2,
to be present at the same time.

One of the interesting results of the previous sections is not only the
fact that one can look at particle exchange in terms of 
markovian processes, but also that the sampling time suitable for
the description of chaotic and turbulent exchange are of the same order of
magnitude for fairly realistic simulations. This suggests the
feasibility of a numerical experiment in which a stochastic term
is added to a time dependent large-scale velocity field.

In addition, in the Introduction we mentioned the issue of
a possible inconsistency of kinematic models as to
the lack of lagrangian conservation of quantities such as
potential vorticity. This difficulty, which has been discussed
at length recently for two-dimensional chaotic flows
(see, above all, Brown and Samelson, 1994;
Balasuriya and Jones, 1996), is overcome in the 
combination of the two mixing processes, as turbulent diffusion
can be seen as a sort of dissipation, which therefore acts
so as to ``smear'' potential vorticity gradients. 

In this numerical experiment we use the model equations 
(\ref{eq:l1}, \ref{eq:l2}) with
${\bf u}^{(M)}({\bf x},t)$ given by the stream function (\ref{eq:phi}) with the
time-dependent perturbation (\ref{eq:pulsaz}) and for the turbulent velocity
$\delta{\bf u}^{(T)}({\bf x},t)$ we use the stochastic process defined in
eqs. (\ref{eq:noise}-\ref{lange}); the parameters are:
$B_{0}=1.2,\;\epsilon=0.3,\;\omega=0.4,\;\theta=\pi/2$
and $\sigma=0.05,\;\tau=T/4$ where $T=2 \pi/\omega$.
As one can see this choice for the parameters 
is simply a superposition of the two previous cases.

Also in this case the matrix elements 
(i.e. the transition probabilities) are 
comparable we the other ones (see Table 1).
As we can see from figs. 15 and 16 the distributions of
the residence times and of the meridional mixing times
display the same qualitative behaviour as the pure diffusive
case (compare with figs. 11 and 13), from which we can 
deduce that for these features the most relevant 
effect is due to the diffusive term.  

As to the correlation function (fig. 17) we note 
that there is a sensible improvement for the markovian
approximation with respect to either the purely chaotic
or the purely diffusive case, since the superposition of the two 
independent perturbations to the large scale flow 
strongly decreases the memory effects. 

\section{Discussion and conclusions}
  
  In this paper particle exchange in a meandering jet 
  has been investigated by means of a kinematic
  model in which mixing is obtained by two different mechanisms:
  chaotic advection and turbulent diffusion.
  The large-scale structure of the jet-like flow is assigned
  in terms of a stationary streamfunction. This has been modified in
  two ways, in order to provide the requested fluid exchange: 
  chaotic advection is induced by adding a time-dependent, relatively 
  small perturbation to the steady portion of the streamfunction.
  Alternatively, turbulent diffusion has been introduced by superimposing
  to this latter a stochastic field. The turbulent field has been selected
  so as to resemble as closely as possible the typical effect of
  upper ocean turbulence in the absence of coherent structures.
  Numerical simulations have been carried out for a case
  in which the two above effects have been jointly present, 
  trying to take into account the richness and complexity of situations
  observed in the ocean, where the two different mixing mechanisms are
  thought to be present simultaneously, even if possibly acting
  at different time and space scales.
  
  The instrinsically different nature of the two investigated mixing
  mechanisms has resulted in the past in disjointed descriptions of 
  their respective effects: chaotic advection in correspondence in
  meandering jets has been studied, e.g., by means of methods derived
  from the dynamical systems theory (Pierrehumbert, 1991; S92; Wiggins, 1992;
  Duan and Wiggins, 1996), 
  whereas the action of turbulent diffusion was clarified
  by phenomenological Lagrangian motion analysis (B91, DGO93).
  
  In this paper mixing is studied in terms of particle transitions among
  areas of the physical two-dimensional space characterized by qualitatively 
  different flow regimes, observed as realizations
  of a markovian process. 
  Given the structure of the velocity field, the partition of the space
  accessible to particles is self-evident and physically consistent.
  A delicate point is obviously the choice of the
  appropriate scale for the sampling time of the process. However, in
  our cases an inherent time scale is present in the velocity field,
  and this is set by, in turn, either the space-time structure of
  the deterministic portion of the flow (chaotic case) or by the
  intrinsic memory time scale present in the stochastic velocity.
  The markovian approach results, in this sense, a quite natural one
  to undertake looking at the overall mixing from a unified perspective,
  embedding elements of dynamical systems and of stochastic process
  theory. Also, it is an alternative way to look at diffusion avoiding the
  usual diffusion coefficients, whose
  general relevance in geophysics has been recently 
  subject to debate (see Artale et al.1997 and Buffoni et al. 1997).
  
  For some fluid exchange properties, 
  the effects of the two above mixing mechanisms 
  are comparable with the results of the markovian 
  approximation: this is the case, for instance,
  of the exit times of particles from the jet and the recirculating regions
  north and south of it. On the other hand, chaotic advection and turbulent
  diffusion act quite differently, under that perspective, when it
  comes to meridional mixing and correlation functions. 
  The failure of
  the markovian approximation for the characterization of the
  meridional particle exchange in the chaotic is due to non-trivial
  long term memory effects.
  Since turbulent diffusion is modelled by a non-white noise process in the
  stochastic velocity field, we would expect for the turbulent case a
  closer behaviour to that predicted by the markovian approximation.
  For the same reason, given the results for the purely chaotic simulations,
  the combined effect of chaos and diffusion has been expected to be well
  described in markovian terms. This is indeed the case, and the 
  results for the joint fluid exchange situations agree quite closely
  with the markovian predictions.
  
  It is worth underlining that
  the possibility to look in terms of a markovian approximation
  at mixing in regions characterized by a quite
  complex flow structure, even in the presence of different transport
  mechanisms, can have quite interesting applicative consequences.
  When the small-scale details of mixing are beyond our interest,
  and if and when our flow system shows fairly well defined time
  scales, it is apparently possible to look at particle exchange
  in a relatively simple manner, over time scales which allow for
  a sensible reduction of the sampling rate. This aspect often turns out to
  be a critical constraint for the undertaking, e.g., of Lagrangian
  investigations of the real ocean, where reducing the required 
  sampling rate can result in reducing the amount of data
  to be collected and transmitted.

\section{Acknowledgements}
We thank V. Artale and M. Falcioni for useful suggestions and 
first reading of the paper. GL would like to thank, also, the 
{\it Istituto di Fisica dell'Atmosfera} of CNR for hosting 
him while working on this paper. 
We are grateful to the ESF Scientific Programme TAO {\it Transport 
Processes in the Atmosphere and the Oceans} for providing meeting
opportunities.
This paper has been partly supported by INFM (Progetto di Ricerca Avanzato
PRA-TURBO) and CNR (Progetto speciale coordinato {\it Variabilit\'a e 
Predicibilit\'a del Clima}).

\section*{Appendix}
\appendix
\section{Some properties of Markov chains}

A Markov chain is a stochastic process in which the
random variable describing the state of the system (in our case
the cell occupied by the tracer)  and the time are discrete
and the probability to be
in a given state at time $n$ depends only on the state at time $n-1$.
All the properties of 
a Markov chain can be derived from the 
transition matrix, $\{W_{ij}\}$, whose elements are the 
probabilities to be in state $j$ at some time $n$ 
being at time $n-1$ in state $i$.
For example the probability to go to state $j$ starting from $i$
in $n$ steps is simply:
\beq
Prob(i\rightarrow j;n)=(W^{n})_{ij}\,.
\label{A0}
\eeq

An excellent introduction to Markov chain can be found
in Feller (1968).
In this appendix we only resume some formulas that
are useful in describing some relevant properties of our system.

First of all we have:
\beq
\sum_{j}W_{ij}=1  \,.
\label{A1}
\eeq
In addition to matrix $\{W_{ij}\}$ one can compute 
the stationary probabilities $P_{i}$ to visit the cells $A_{i}$ as 
elements of the 
(left) eigenvector corresponding to the eigenvalue 1 :
\beq
P_{j}=\sum_{i}P_{i}W_{ij}\,,
\label{A2}
\eeq
Let us note that eqs. (\ref{A1},\ref{A2})  are rather general
results that hold for a generic discrete stochastic process. 
Eq. (\ref{A1}) describes the conservation of probability and 
eq. (\ref{A2}) is nothing but Bayes' Theorem.
In order to have ergodicity and mixing properties the 
Markov chain must have a non zero probability to 
pass through any state in a finite number of steps, i.e.
there exist a $n$ such that $(W^{n})_{ij} >0$ (Feller 1968).
Defining $\rho_{i}(t)$ the probabilities to visit 
the state $i$ at time $t$, for a Markov chain we have:
\beq
\rho_{j}(t+1)=\sum_{i}\rho_{i}(t)W_{ij}
\label{A3}
\eeq
in this way we have the evolution of the probability vector 
$\rho_{i}$ (see Rickman and Froeschl\`e, 1979),
eq. (\ref{A2}) corresponds to $t\rightarrow \infty$ of eq. (\ref{A3}), i.e.
the equilibrium distribution $\rho_{i}(\infty)=P_i$.

The probability of the first exit times from 
a state $i$ can be simply defined as
the probability to stay for $n-1$ steps in state $i$ times the 
probability to exit at step $n$, i.e.:
\beq
P_{i}(n)=(1-W_{ii}) (W_{ii})^{n-1}\,.
\label{A4}
\eeq
that can be rewritten as:
\beq
P_{i}(n)=\left[ \frac{(1-W_{ii})}{W_{ii}}\right] \exp(-n\alpha)\,,\; 
{\mbox {with}} \;
\alpha=|\ln W_{ii}|\,. 
\label{A5}
\eeq

In a similar way we can define the probability
$f_{ij}(n)$  of the first arrival from state $i$ to state $j$ 
at step $n$. This is the probability to arrive to state 
$j$ starting from $i$ in $n$ step i.e. $(W^{n})_{ij}$ 
minus the probability of first
arrival at step $n-k$ times the probability of return in $k$ steps,
i.e. $(W^{k})_{jj}$ with $k=1,\cdots,n-1$: 
\beq
f_{ij}(n)=(W^{n})_{ij}-\sum_{k=1}^{n-1} f_{ij}(n-k) (W^{k})_{jj}.
\label{A6}
\eeq

For each state of a Markov process a correlation function can be defined
for the variable $\chi_i(n)$ which is equal to 1 if at time
$n$ the $i$ state is visited and to zero otherwise
(see eqs. \ref{eq:chi}-\ref{eq:corr}).
The normalized correlation function
\beq
C_{i}(n)\,=\,\frac{<\chi_i(0)\chi_i(n)> - <\chi_i(0)>^2}
{<\chi_i(0)^2> - <\chi_i(0)>^2}.
\label{A7}
\eeq 
is strictly related to the diagonal element $(W^n)_{ii}$ of the 
transition matrix to the n-th power and to the stationary frequency $P_i$. 
Notice that 
\beq
<\chi_i(0)>\, = \, P_i\, ,  \;\;
<\chi_i(0)^2>\,=\,P_i .
\label{A8}
\eeq

Furthermore, being $P_i$ the probability that the 
initial state at $n=0$ be $i$ and $(W^n)_{ii}$ the probability
to be in $i$ again after $n$ iterations, one has:
\beq
<\chi_{i}(0)\chi_{i}(n)> \, = \, P_i \cdot (W^n)_{ii}
\label{A9}
\eeq
and therefore:
\beq
C_{i}(n)= ( (W^n)_{ii} - P_i ) / (1 - P_i))
\label{A10}
\eeq

\newpage
\centerline {TABLES}
\vspace{1cm}
Table 1

Transition matrix elements
\vspace{.5cm}

\begin{tabular}{|l|l|l|l|}\hline 
 $W_{ij}$ & Case A & Case B & Case C \\ \hline \hline
 $W_{11}$ & .66 & .74 & .58\\ \hline
 $W_{12}$ & .17 & .13 & .21 \\ \hline  
 $W_{13}$ & .17 & .13 & .21\\ \hline  
 $W_{21}$ & .12 & .09 & .14\\ \hline  
 $W_{22}$ & .88 & .91 & .86\\ \hline  
 $W_{23}$ & .00 & .00 & .00\\ \hline  
 $W_{31}$ & .12 & .09 & .14\\ \hline  
 $W_{32}$ & .00 & .00 & .00\\ \hline 
 $W_{33}$ & .88 & .91 & .86\\ \hline
\end{tabular}

\vspace{1cm}
Table 2

Visit probabilities
\vspace{.5cm}

\begin{tabular}{|l|l|l|l|}\hline 
 $P_{i}$ & Case A & Case B & Case C\\ \hline \hline
 $P_{1}$ & .26 & .25 & .25\\ \hline
 $P_{2}$ & .37 & .375 & .375 \\ \hline  
 $P_{3}$ & .37 & .375 & .375\\ \hline  
\end{tabular}

\vspace{1cm}

Case A -  Deterministic chaotic model defined by eq. (\ref{eq:hamil})
related to the stream function (\ref{eq:phi})
with parameters $L=7.5,\;B_0=1.2,\;c=0.12,\;\omega=0.4,\;
\epsilon=0.3$.

Case B -  Turbulent diffusion model defined by eqs. 
(\ref{eq:noise}-\ref{lange}) with parameters 
$\sigma=.05$ $\tau=T/4$.

Case C - Model with chaotic advection plus turbulent diffusion
with the same parameters of case A and B.

The statistics have been computed over $2\,10^{6}$ periods.

\newpage

\centerline{FIGURE CAPTIONS}

\begin{itemize}

\item FIGURE 1: Snapshot of the velocity field derived from the stream
function eq.(\ref{eq:phi}) with $L=7.5,\;B_{0}=1.2,\;c=0.12$.

\item FIGURE 2: Stream lines of the time-dependent
stream function eq.(\ref{eq:phi}), with $B$ given by 
eq.(\ref{eq:pulsaz}), $B_0=1.2$,
$\omega=0.4$ and $\epsilon=0.3$ ($T=2\pi/\omega$), at 
three different times:
(a) $t=0. \,\, (T/2)$, (b) $t=T/4$ and (c) $t=3T/4$.

\item FIGURE 3: Critical values of the periodic 
perturbation amplitude for the overlap of the resonances, 
$\epsilon_{c}/B_{0}$ vs $\omega/\omega_{0}$, for the 
stream function (\ref{eq:phi}) with $L=7.5,\;B_{0}=1.2,\;c=0.12$ and
$\omega_{0}=.25$, which is the typcal frequency for the rotation 
of a tracer on the boundary of the recirculation gyres.  
The critical values have been estimated following a cloud of $100$
particles initially located between the states $1$ and $2$ up to
$500$ periods.

\item FIGURE 4:
Spreading of a cloud of $5000$ tracers at different times
for the deterministic model (see Fig. 2):
(a) $t=0$ all tracers are inside a very small square,
(b) $t=T$ ,(c) $t=5\,T$, (d) $t=10\,T$, (e) $t=20\,T$ and (e) 
$t=100\,T$. 

\item FIGURE 5: Probability distribution 
of the first exit times from states 1 (diamonds), 2 (squares) 
and 3 (crosses) for the deterministic model (see Fig. 2). 
The straight lines are the markovian predictions given by 
eq.(\ref{A4}) with $W_{ii}$ of table 1 (case A). 
The time unit is the period $T$ of the perturbation. 
The statistics is computed over $2 \,10^6$ periods. 
 
\item FIGURE 6: Correlation functions of the states 1 (diamonds) and
 2 (crosses)  compared with the
markovian predictions (continuous lines) eq. (\ref{A10}) for the deterministic 
model (see Fig. 2).

\item FIGURE 7: Probability distribution 
of the meridonal mixing times (MMT)
compared with the markovian predictions (continuous lines) eq. (\ref{A4})
for the deterministic model (see Fig. 2).

\item FIGURE 8: Comparison beetwen the symbolic 
sequence of the states as function of time (for $t=T,2T,\cdots$) 
obtained by the integration of the deterministic model 
equations (1,2) (dotted lines)
and the symbolic sequence generated from the Markov chain (solid lines)
defined by the transition matrix computed 
as described in eq.(\ref{eq:matrix}) and reported in table 1 (case A) .  
Here $0$ represents the state 1 i.e. the jet 
while $1,-1$ the recirculation gyres i.e. the states 2,3.
  
\item FIGURE 9: Block entropies $h_n$ vs $n$ (\ref{eq:block}) 
for the deterministic model (see Fig. 2), computed from
a sequence of $10^6$ symbols.

\item FIGURE 10: The same as figure 4 for the stochastic model 
given by eqs.(\ref{eq:noise},\ref{lange}), 
with parameters $\sigma=0.05$ and $\tau=T/4$. 
The time unit $T$ is set equal to the period of the deterministic 
perturbation (see Fig. 2).
 
\item FIGURE 11:  Probability distribution 
of the first exit times from states 1,2 and 3 
for the stochastic model (see Fig. 10). 
The straight lines are the markovian predictions given by 
eq.(\ref{A4}) with $W_{ii}$ of table 1 (case B).
 
\item FIGURE 12:  Correlation functions compared with the
markovian predictions (continuous lines) eq. (\ref{A10}) for the 
stochastic model (see Fig. 10).

\item FIGURE 13: Probability distribution 
of the MMT compared with the markovian predictions 
 eq. (\ref{A6}) for the stochastic model (see Fig. 10).
 
\item FIGURE 14: Block entropies $h_n$ vs $n$
for the stochastic model, computed 
from a sequence of $10^6$ symbols.

\item FIGURE 15: Probability distributions 
of the first exit times from states 1,2 and 3 
in the model with chaotic advection combined with turbulent diffusion
(sect. 4.3) with parameters:
$B_0=1.2$, $\omega=0.4$, $\epsilon=0.3$  
and $\sigma=0.05$, $\tau=T/4$ where $T=2\pi/\omega$. 
The straight lines are the markovian predictions 
given by eq. (\ref{A4}) with $W_{ii}$
of table 1 (case C). 
 
\item FIGURE 16:  Probability distributions 
of the MMT compared with the markovian predictions 
 eq.(\ref{A6}) for the model 
and the parameters of figure 15.

\item FIGURE 17:  Correlation functions
compared with the markovian predictions  eq.(\ref{A10}) 
for the model and the parameters of figure 15.

\end{itemize}

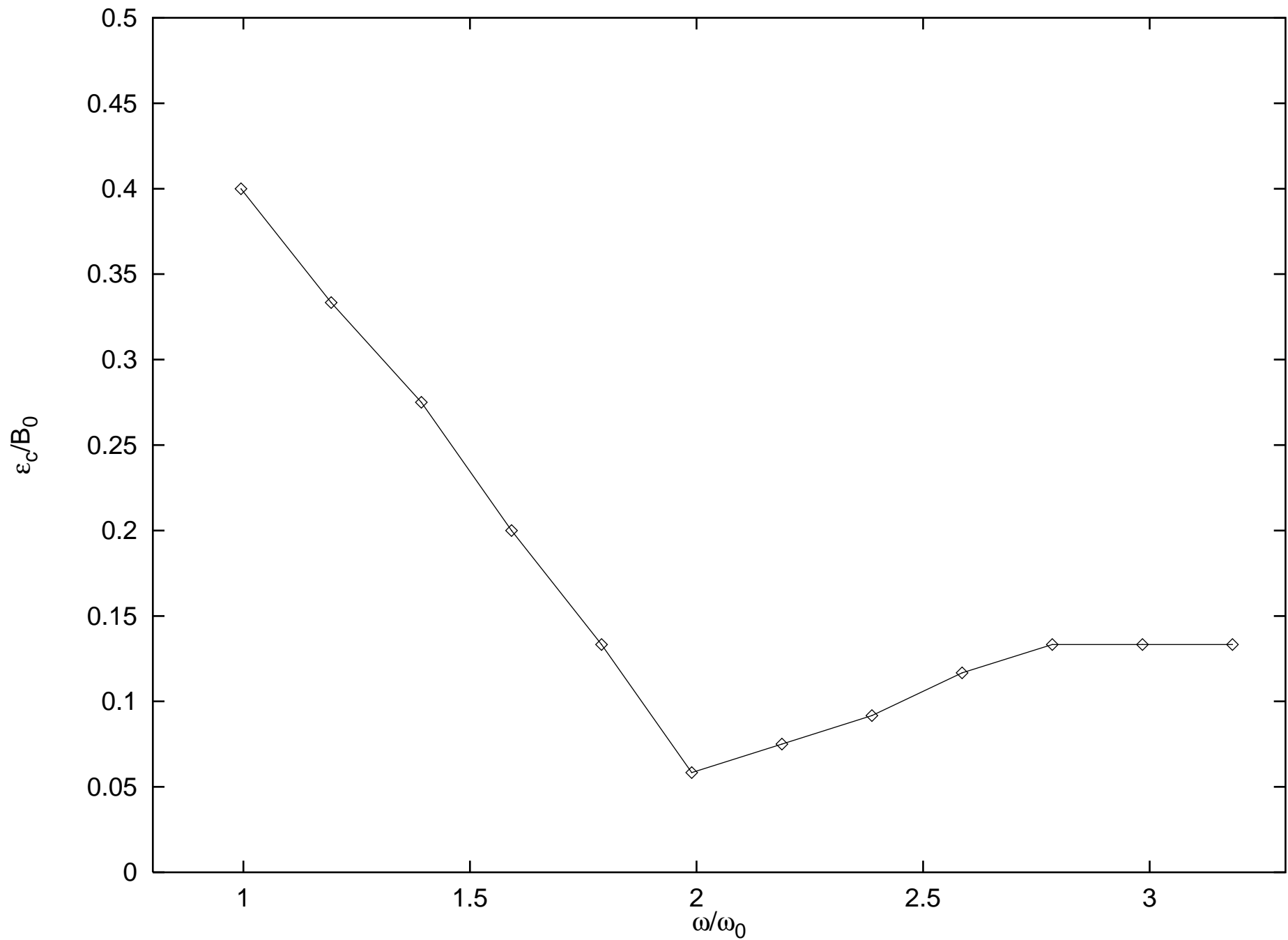
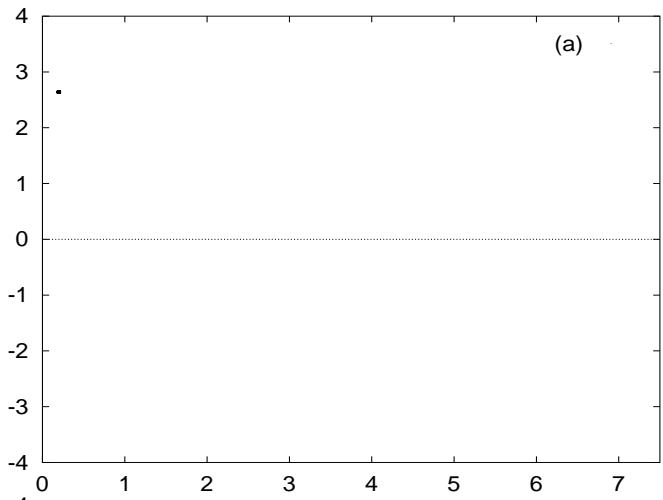
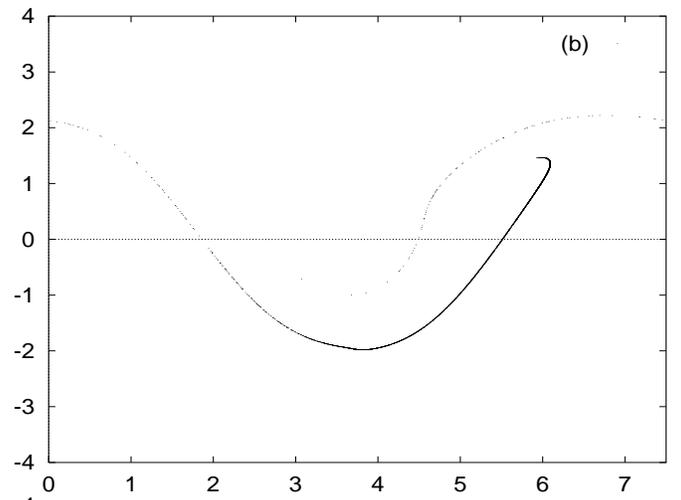
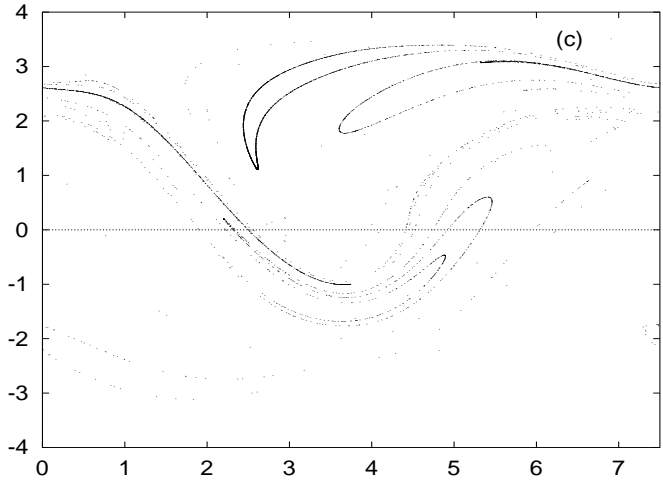
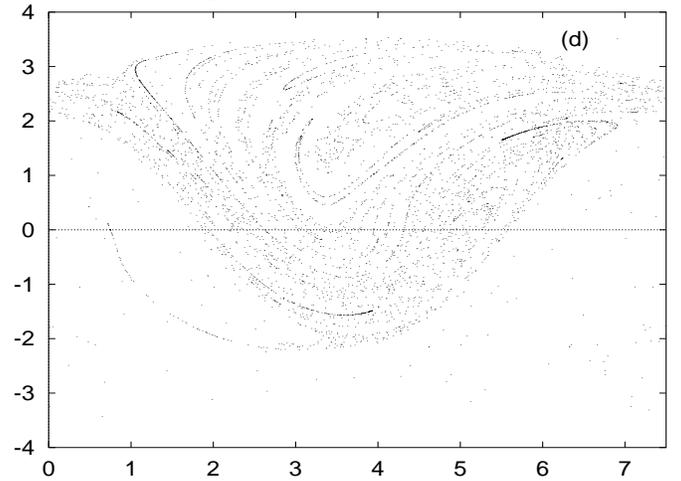
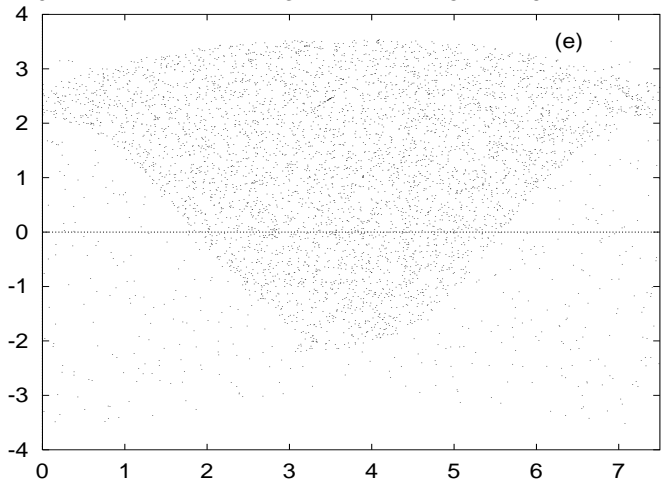
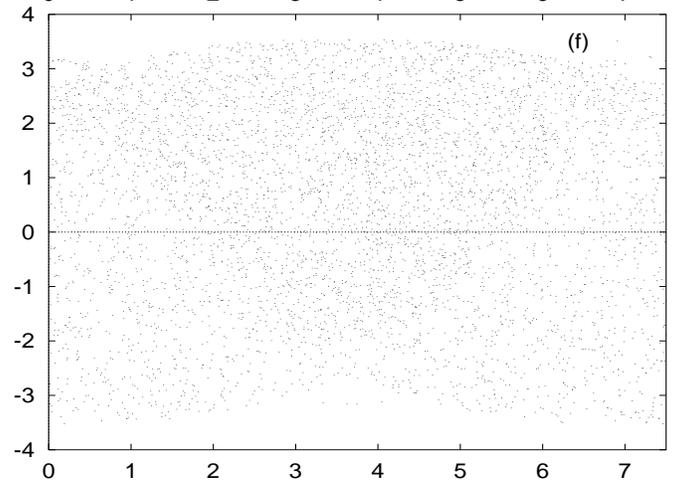
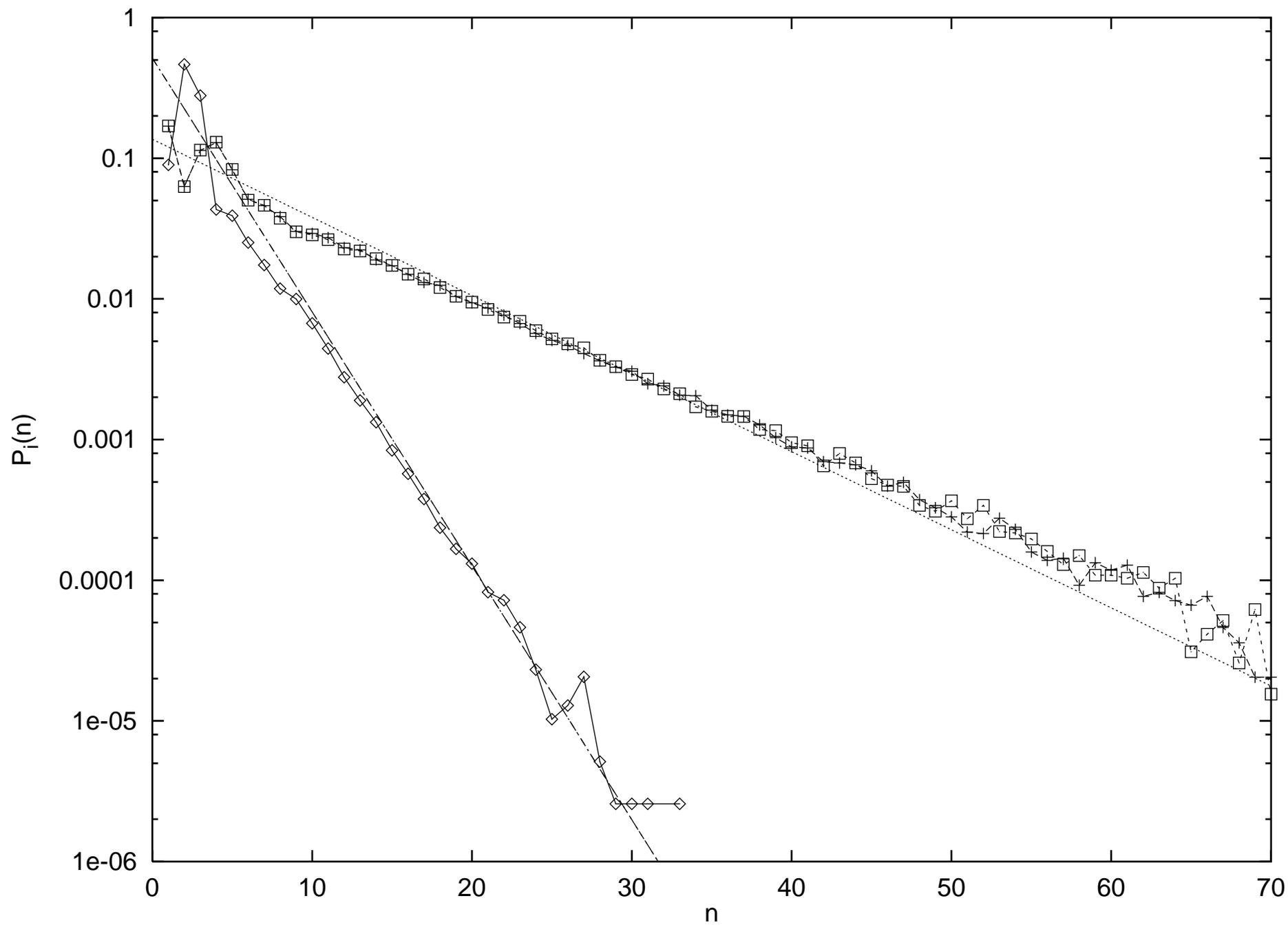
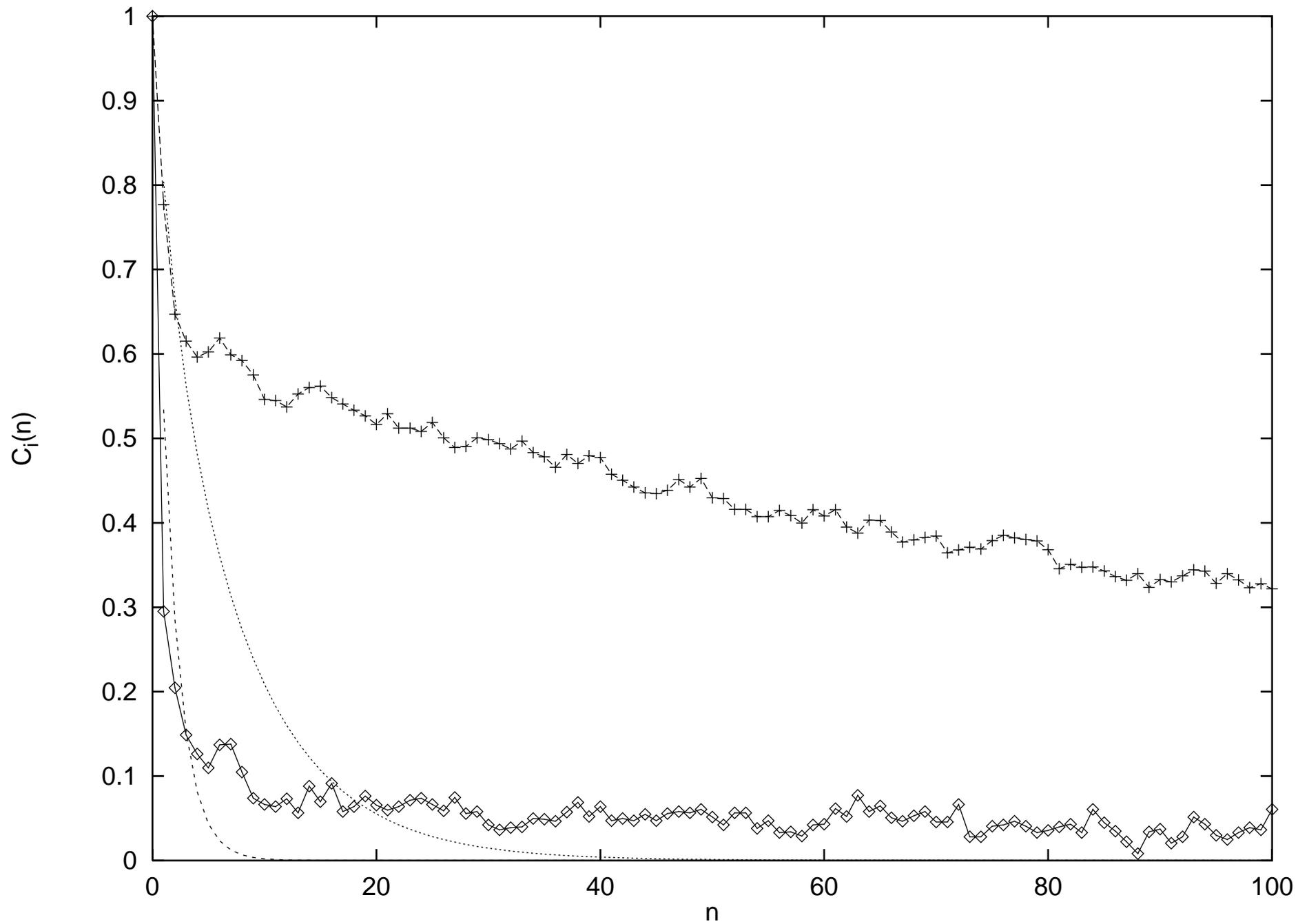
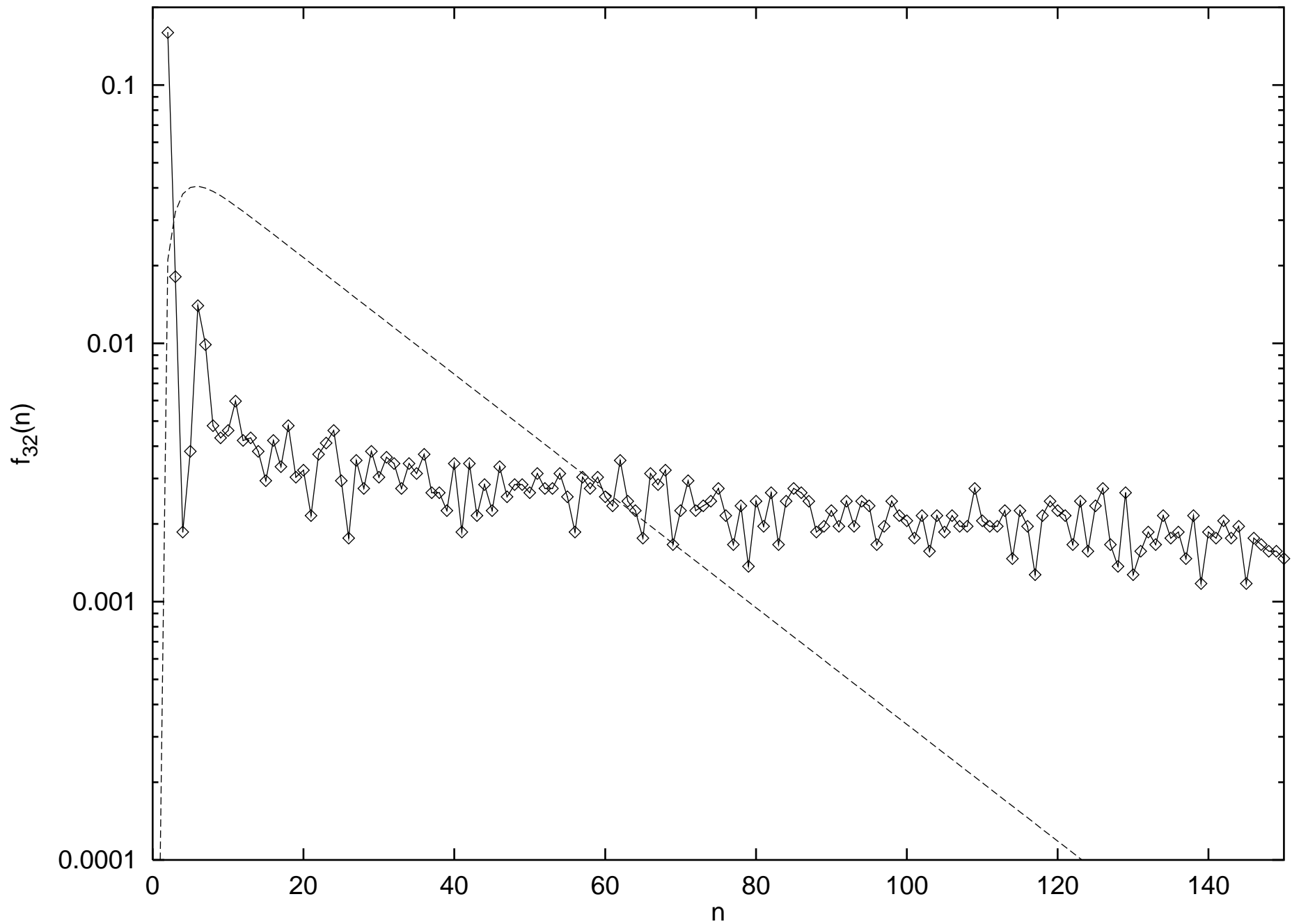
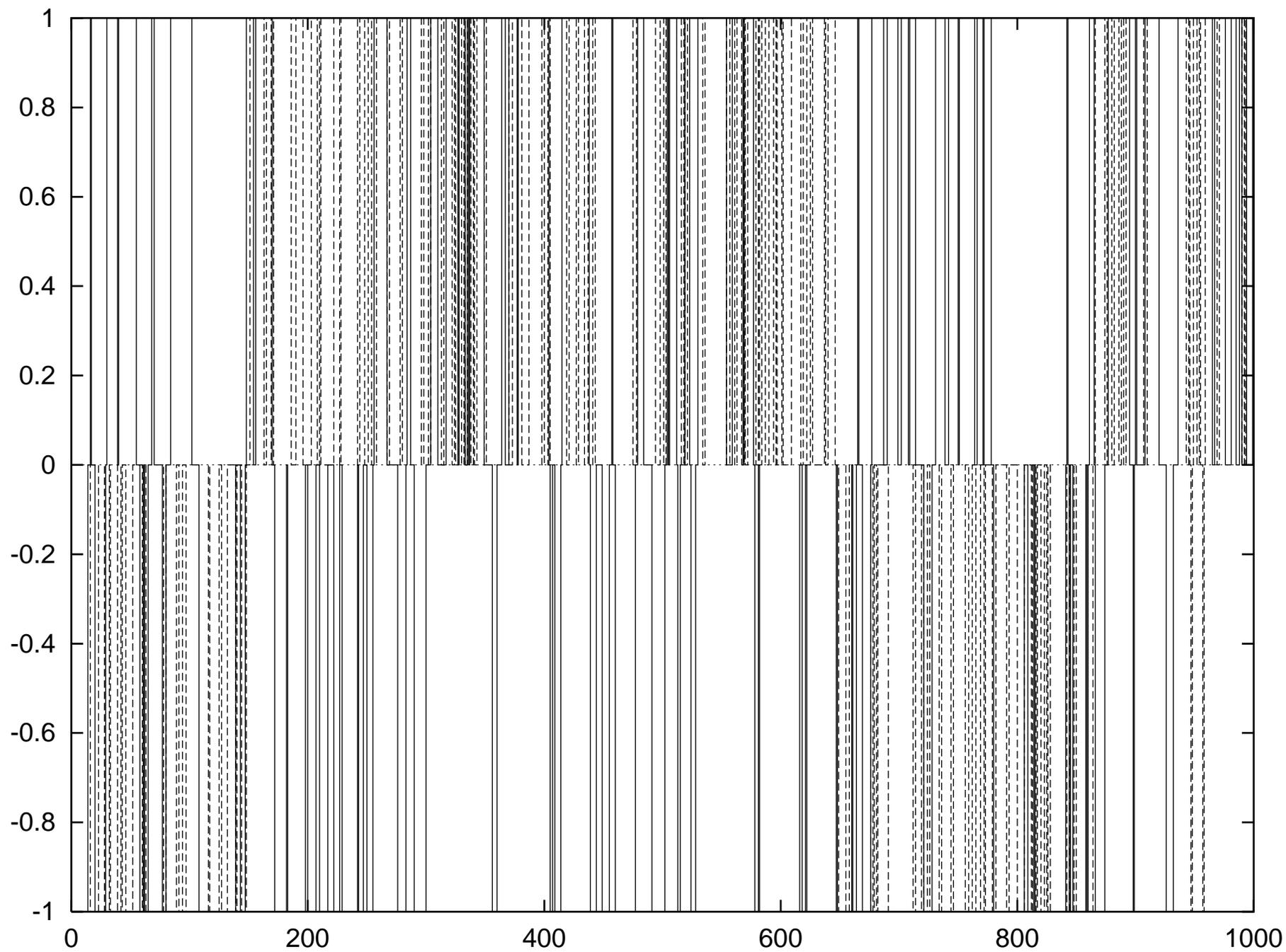
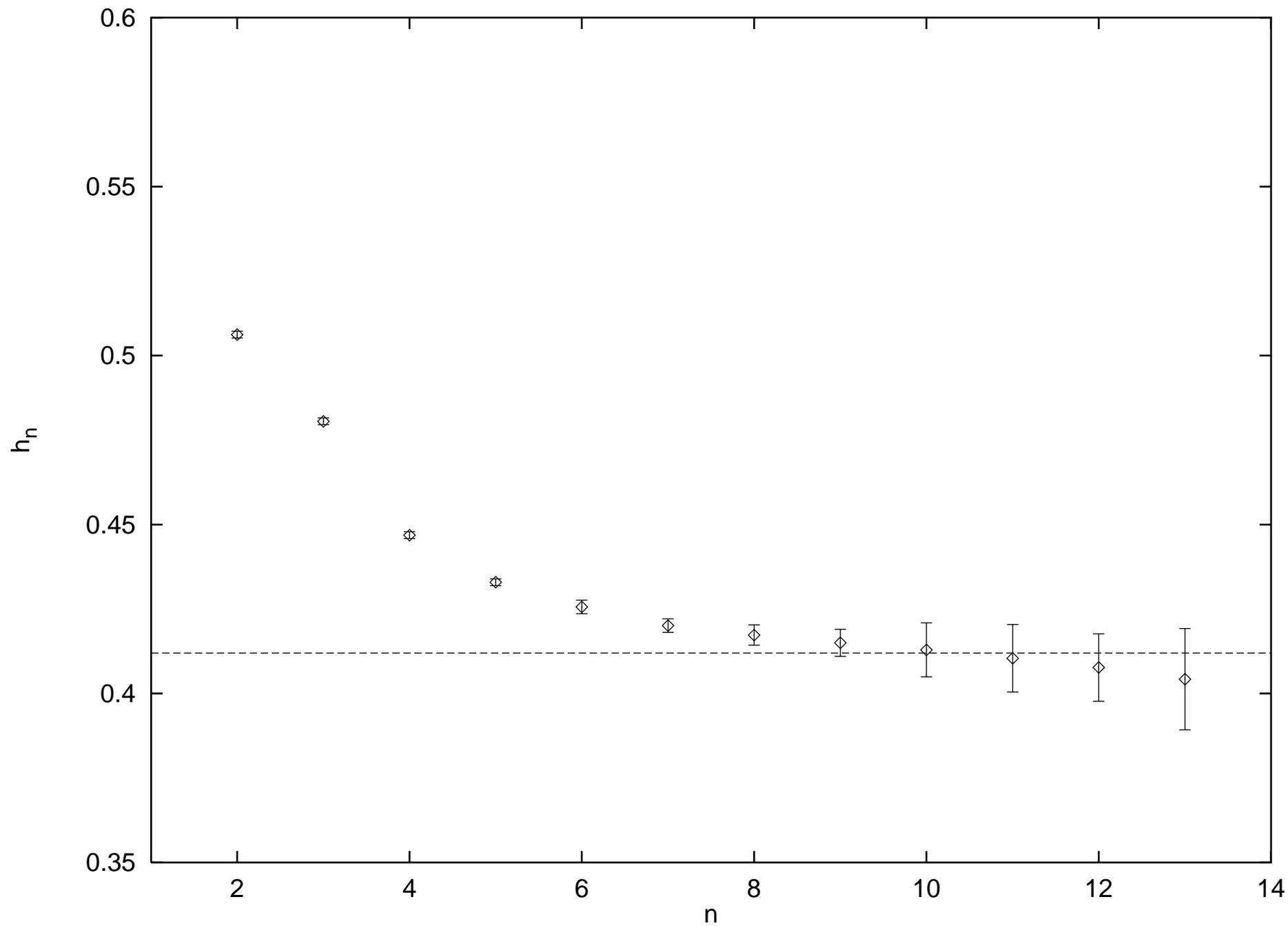
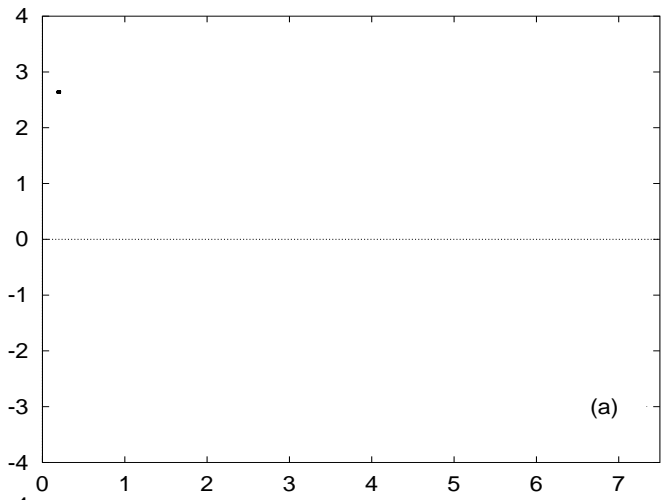
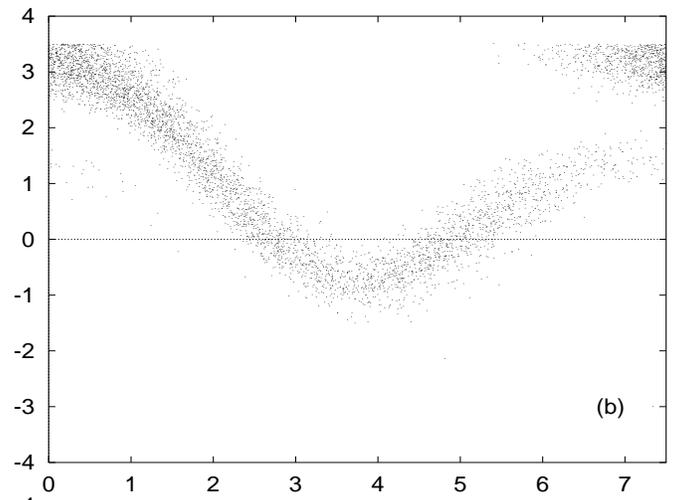
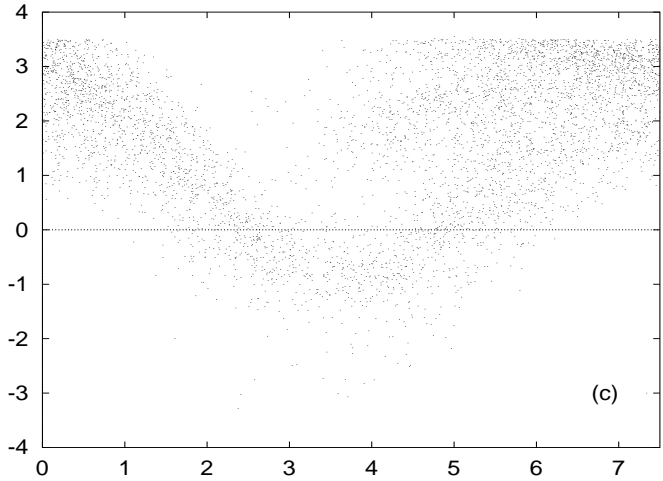
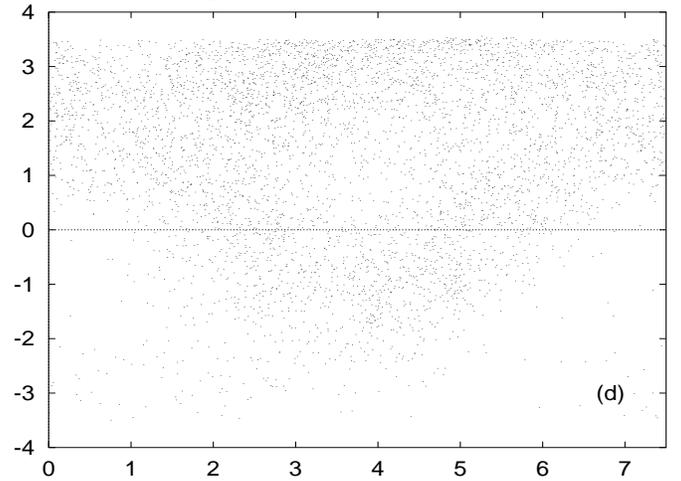
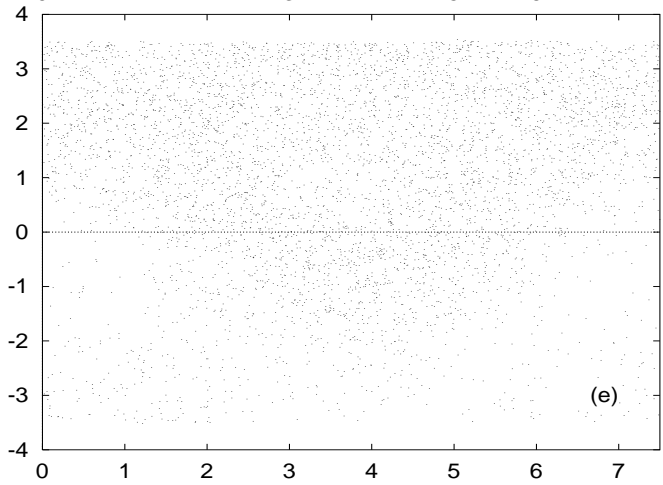
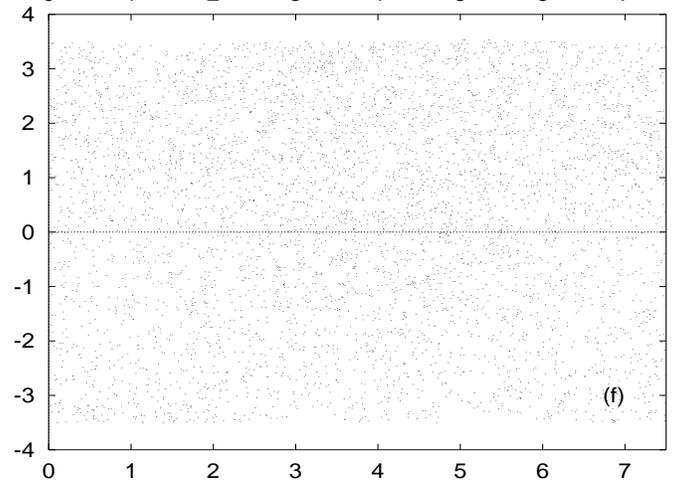
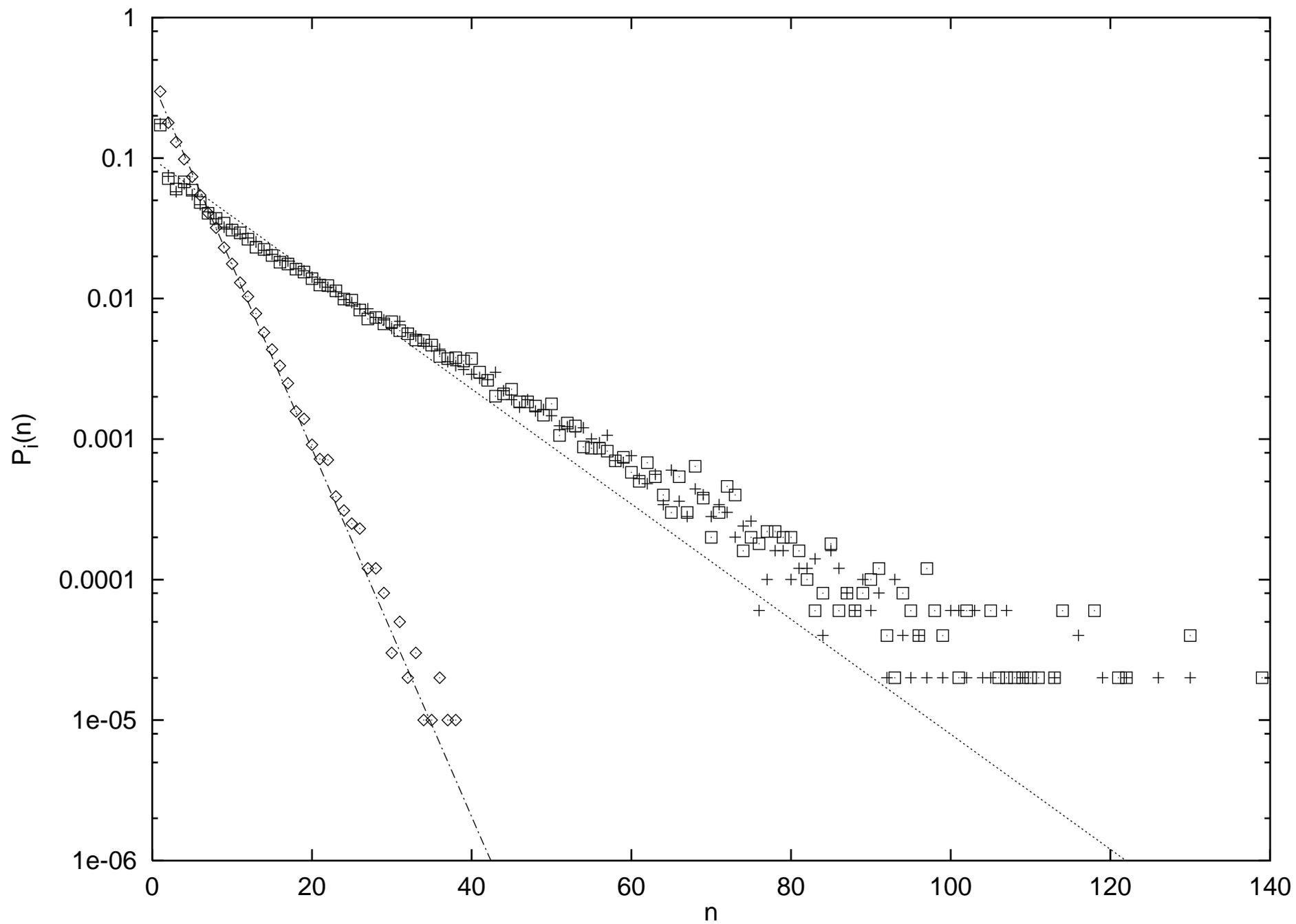
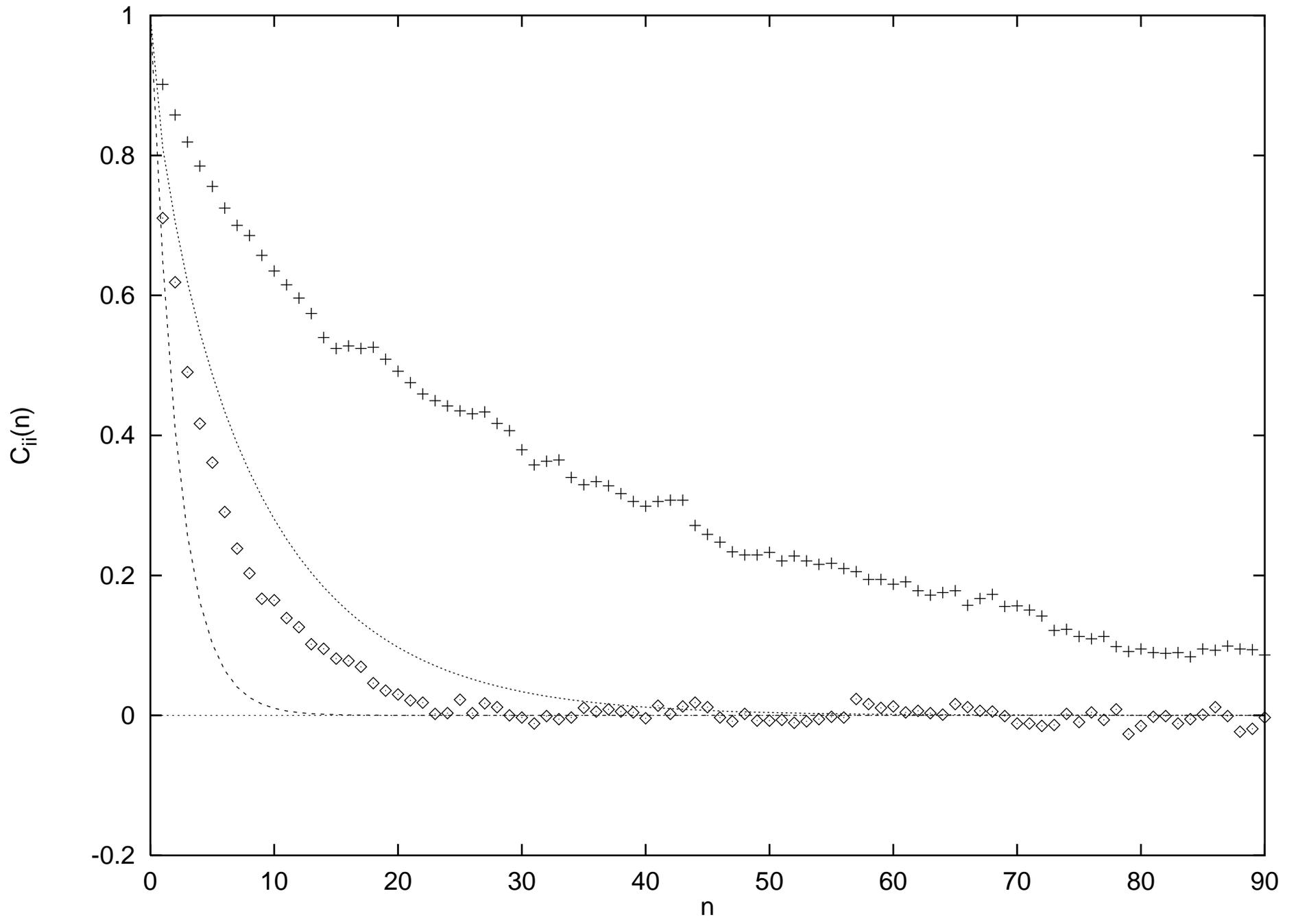
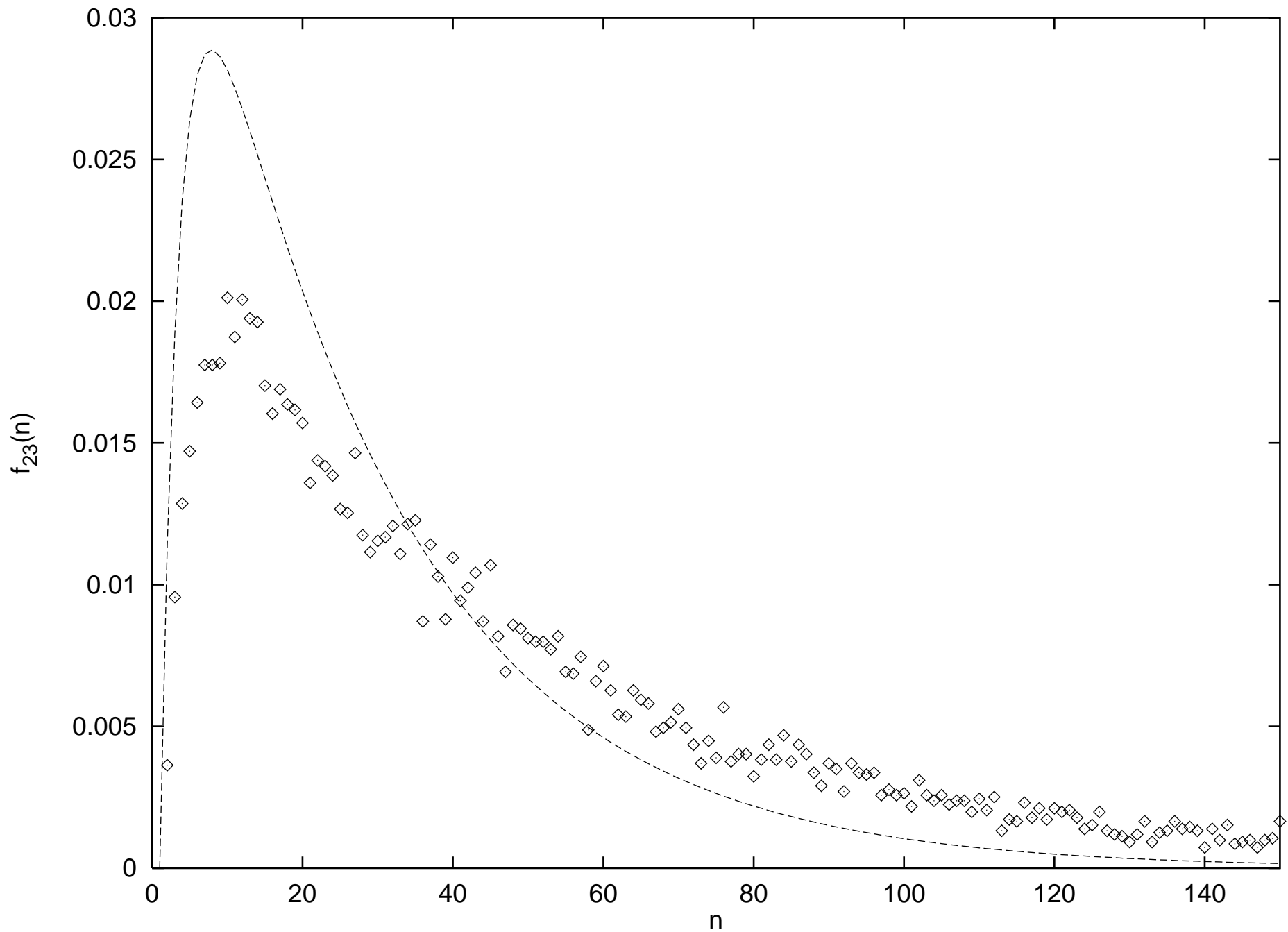
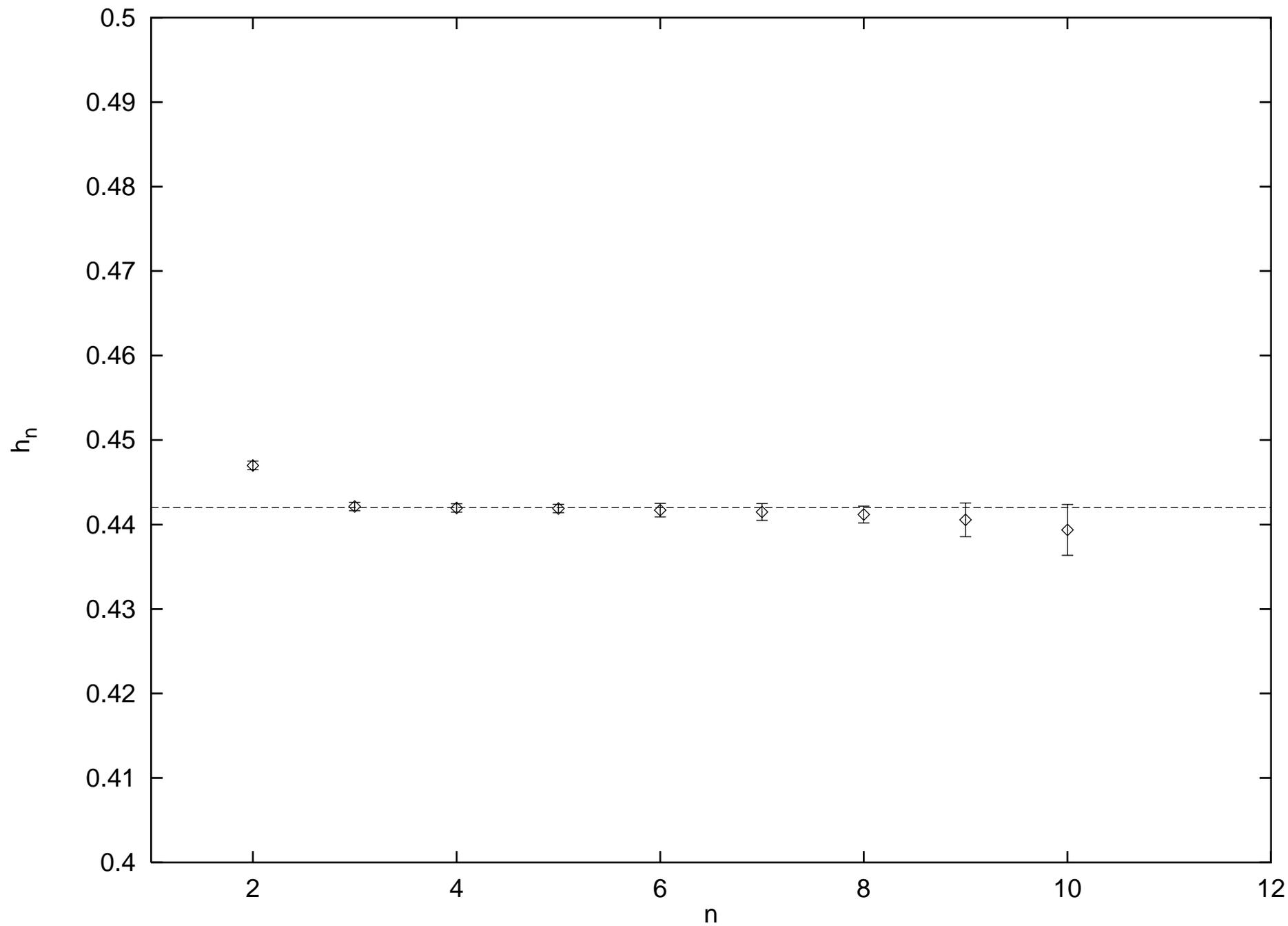
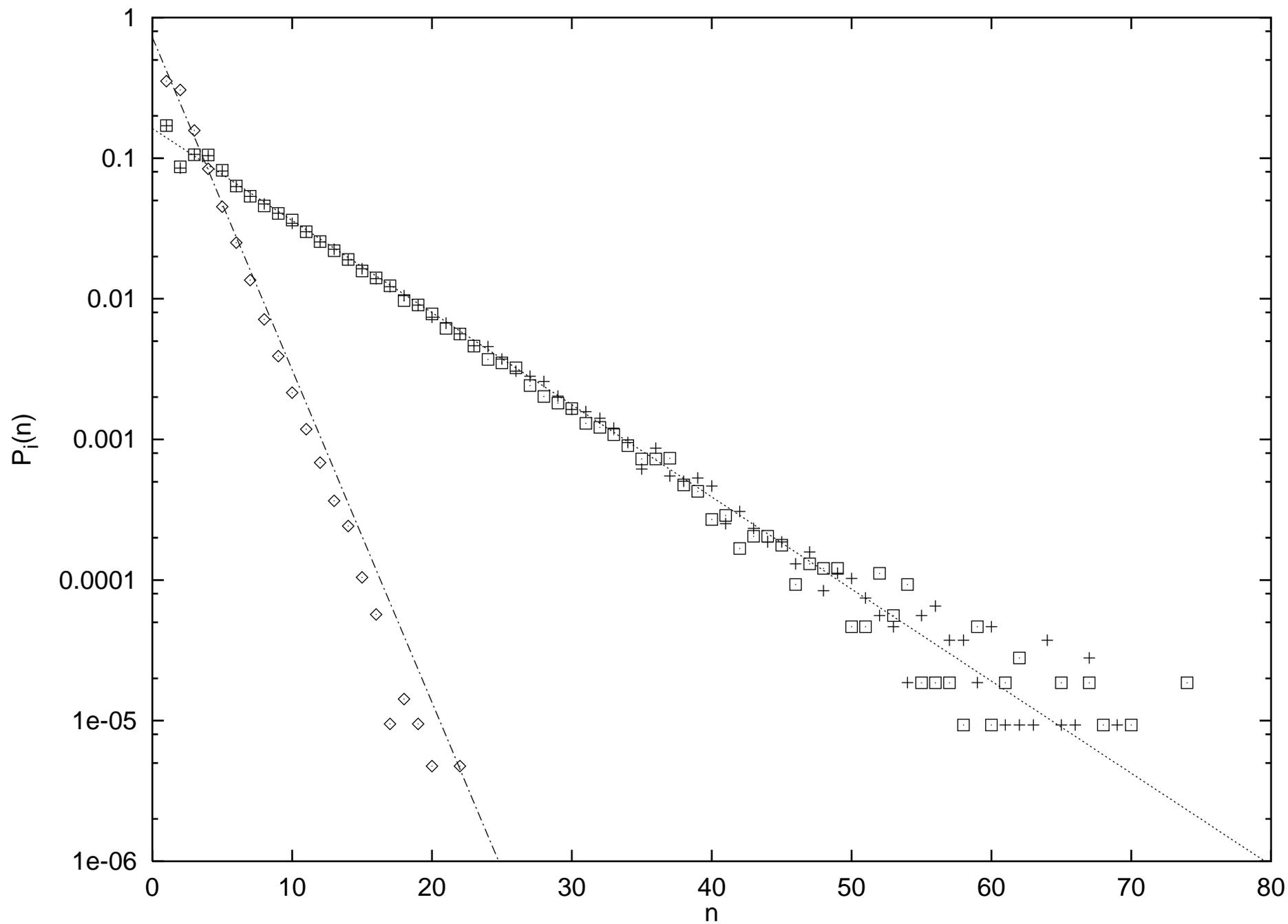
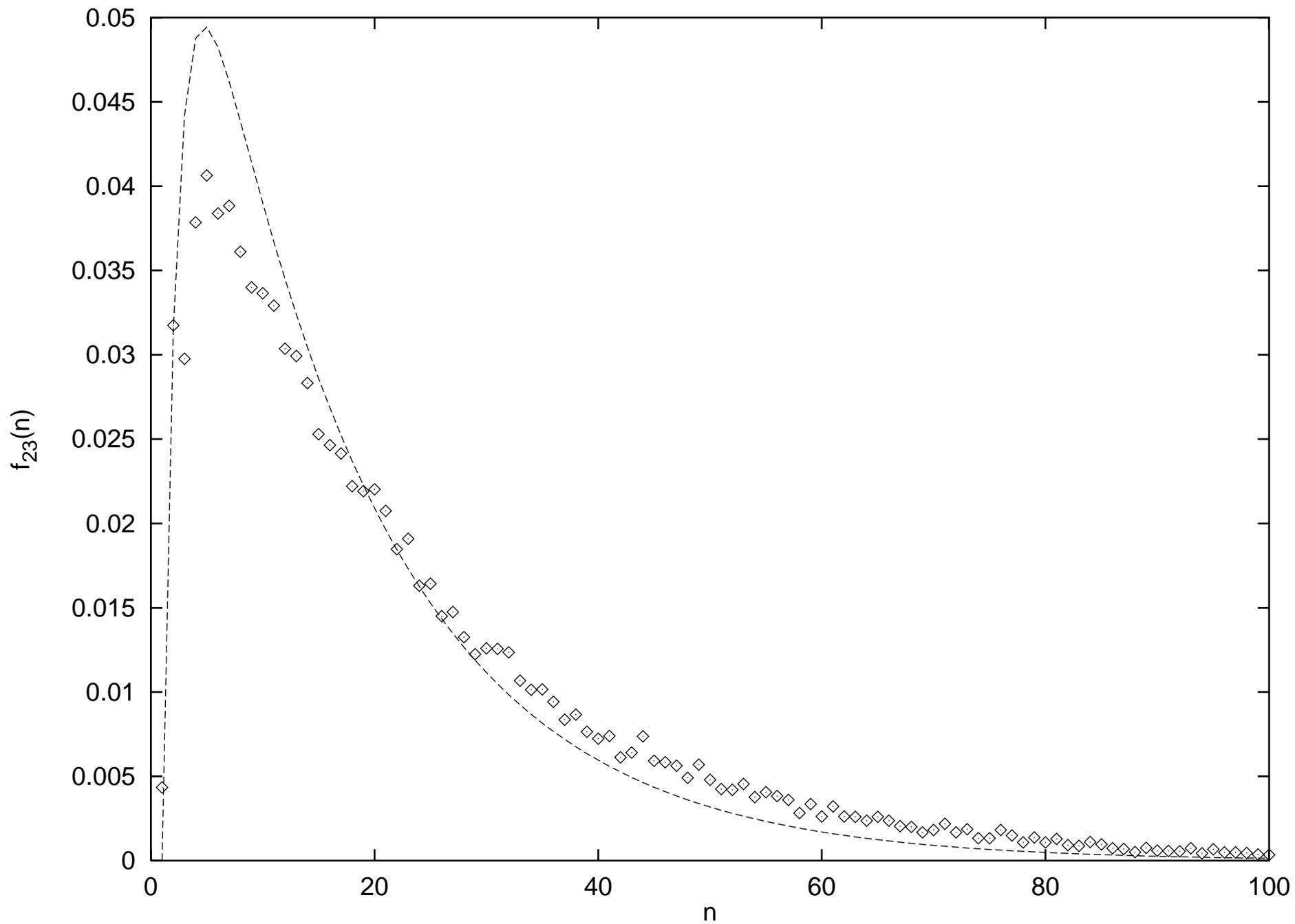
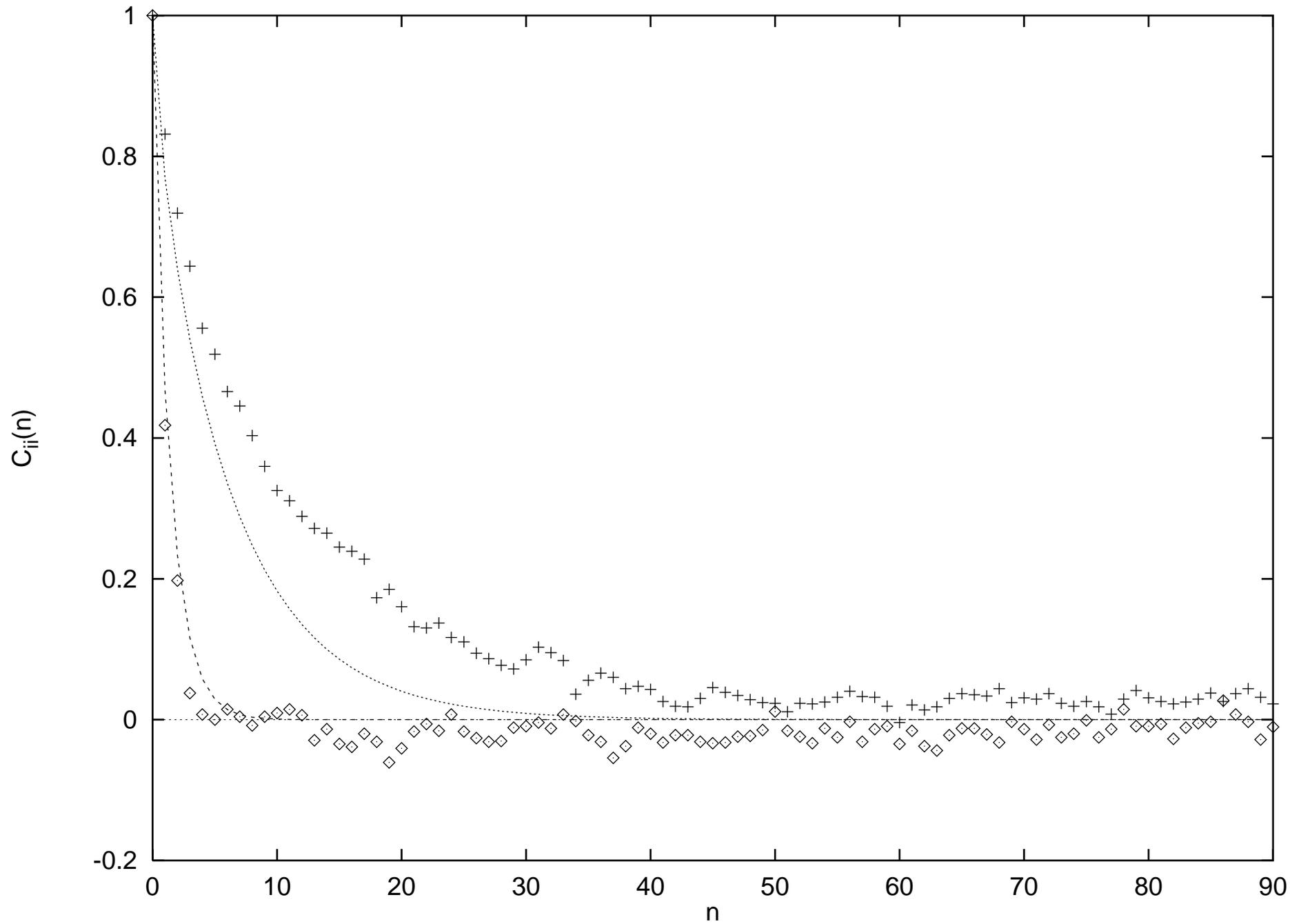
\end{document}